\documentclass[11pt, letterpaper]{article}
\pdfoutput=1
\usepackage{amssymb,amsmath,amsthm}
\usepackage{hyperref}
\usepackage{graphicx,color}

\usepackage{appendix}

\usepackage[utf8]{inputenc}

\usepackage{microtype}

\numberwithin{equation}{section}

\newtheorem{thm}{Theorem}
\newtheorem{rem}{Remark}
\newcommand{\uD}{D}
\newcommand{\ud}{d}

\newcommand{\ie}{{\it i.e.}}
\newcommand{\eg}{{\it e.g.}}

\newcommand{\LB}[3][\Lambda]{[\, #2  \,_{#1}\,  #3 \,]}

\newcommand{\LP}[3][\Lambda]{ [ \, #2  \,_{#1}\,  #3 \,  ] }
\newcommand{\M}{\mathcal{M}}
\newcommand{\G}{\mathcal{G}}
\newcommand{\Dp}{D_{\!+}}
\newcommand{\Dm}{D_{\!-}}

\hypersetup{
    pdftitle={Sheaves of $N=2$ supersymmetric vertex algebras  on Poisson manifolds},    
    pdfauthor={  Joel Ekstrand, Reimundo Heluani and Maxim Zabzine},     
    pdfkeywords={Vertex algebra, SUSY Vertex algebra, Poisson vertex algebra, Poisson geometry},
    pdfnewwindow=true,      
    colorlinks=true,       
    linkcolor=black,          
    citecolor= black,        
    filecolor= black,      
    urlcolor= black           
}

\newcommand{\be}{\begin{equation}} 
\newcommand{\ee}{\end{equation}} 
\newcommand{\bea}{\begin{eqnarray}} 
\newcommand{\eea}{\end{eqnarray}}

\begin{document}

\thispagestyle{empty}
\begin{flushright} \small
UUITP-24/11  \\
NSF-KITP-11-195\\
 \end{flushright}
\smallskip
\begin{center} \LARGE
{\bf  Sheaves of $N=2$ supersymmetric vertex algebras  on Poisson manifolds}
  \\[12mm] \normalsize
{\bf Joel~Ekstrand$^a$, Reimundo~Heluani$^b$ and Maxim Zabzine$^{a,c}$} \\[8mm]
 {\small\it
 $^a$Department of Physics and Astronomy, 
     Uppsala university,\\
     Box 516, 
     SE-751\;20 Uppsala,
     Sweden\\
     ~\\
     $^b$ IMPA, Rio de Janeiro,
     RJ 22460-320, Brazil\\
     ~\\
       ${}^c$Kavli Institute for Theoretical Physics, University of California, \\
Santa Barbara, CA 93106 USA\\
}
\end{center}
\medskip
\begin{abstract}
 \noindent  
 We construct a sheaf of  $N=2$ vertex algebras naturally associated to any Poisson manifold. The relation of this sheaf to the 
 chiral de Rham complex is discussed.  We reprove the result  about the existence of two commuting $N = 2$ superconformal 
  structures on the space of sections of the chiral de Rham complex   of a Calabi-Yau manifold, but now calculated in a manifest $N=2$
    formalism.    We discuss how the semi-classical limit of  this sheaf of  $N=2$ vertex algebras is related to the classical 
    supersymmetric non-linear sigma model.
\end{abstract}   

\medskip

\section{Introduction}

To any smooth manifold $M$ one can associate a sheaf of vertex algebras \cite{Malikov1999}, which is called the chiral de Rham 
 complex (CDR) of $M$. Locally on a $d$-dimensional manifold $M$ one attaches $d$ copies of the free bosonic $\beta\gamma$-system tensored with $d$ copies of the free fermionic $bc$-system. These local models are then glued along intersections of the corresponding patches on $M$ using appropriate automorphisms of these free field systems.  One can combine these $4d$ fields into $2d$ $N=1$ \emph{superfields}  
 to obtain a sheaf of  $N=1$ SUSY vertex algebras \cite{Ben-Zvi2008}.  
  More generally, one can construct a sheaf of $N=1$ SUSY vertex algebras associated to any Courant algebroid $E$  over $M$, 
   and this can be done in  a coordinate free fashion \cite{Heluani2008a}. Geometric properties of $M$ are reflected in algebraic properties of CDR. For example,  if $E$ admits a generalized Calabi-Yau structure, then there exists an embedding of the $N=2$ superconformal 
      vertex algebra into global sections of this sheaf \cite{Heluani:2008hw}. The reader may find more results along these lines in \cite{Ben-Zvi2008, Heluani2008a,
       Ekstrand2010a}.  
  
  There is a quasiclassical version of CDR as a sheaf of Poisson vertex algebras \cite{malikov-lagrangian}. 
   This can be naturally related to the Hamiltonian treatment of supersymmetric classical non-linear sigma models \cite{Ekstrand2009c}.
  Indeed, CDR can be interpreted as a formal quantization of the non-linear sigma model. By ``formal'' we mean here that instead of working with the actual loop space of $M$, one deals with the space of formal loops into $M$ \cite{kapranov}. Nevertheless, the relation 
    to sigma models is quite inspiring, e.g. see \cite{Ekstrand2010a}.  
 
 In this  note, we present a very simple construction of a sheaf of  $N=2$ SUSY vertex algebras on any  Poisson manifold $M$. 
  We discuss the relation of this $N=2$ sheaf to CDR as a sheaf of  $N=1$ SUSY vertex algebras. 
  We recover the main result from \cite{Heluani2008a}, about the existence of two 
commuting $N = 2$ superconformal structures on the space of sections of 
CDR in the Calabi-Yau case, but now calculated 
in a manifest $N = 2$ formalism. We also briefly discuss  the semiclassical limit of this $N=2$ sheaf and its relation to the Hamiltonian 
 treatment of $N=(2,2)$ supersymmetric sigma models with a K\"ahler target.
  
The paper is organized as follows. In Sect.\ \ref{secVAdef} the definition of  an $N_K=2$ SUSY Vertex algebra is given.  
 In Sect.\ \ref{s-Def-N2sheaf} we construct the sheaf of N=2 SUSY vertex algebras on any Poisson manifold.  
  In Sect.\ \ref{s-symplectic} we discuss the case of a symplectic manifold.
  Sect.\ \ref{s-CY} deals with the case of a Calabi-Yau manifold.
 In  Sect.\ \ref{sec:LagtoHam} we consider the semiclassical limit of the $N=2$ sheaf and discuss the relation to 
  the $N=(2,2)$ supersymmetric sigma model. 
  Section \ref{s-summary} contains a summary of the results in this article and a discussion of open questions. 
 All technical calculations are collected in the appendices.
For the reader's convenience,  we collect  the rules for $\Lambda$-brackets  in Appendix \ref{app:rulesLambdabr}. 
  Appendix \ref{proofGdiagN2alg} contains the calculation for the symplectic case.  Appendix \ref{proofGpGmN2alg}
   presents the proof of the existence of an embedding of the  $N=(2,2)$ superconformal algebra in the Calabi-Yau case, in manifest $N=2$ formalism. 
     Appendix \ref{app:hamderivation} contains the details of the Hamiltonian treatment of the $N=(2,2)$ sigma model with a
      K\"ahler target.

\section{$N=2$ SUSY Vertex algebra}\label{secVAdef}
In this section we briefly review the definitions of vertex algebras and their $N=2$ supersymmetric counterparts. The number of supersymmetries introduced are in general arbitrary, but since we are mainly interested in the case of two supersymmetries in this work, we choose to be concrete. For more details, the reader is referred to \cite{Kac1998} and \cite{Heluani2007}. 

Given a vector space $V$, a \textit{field} is defined as an $\text{End}(V)$-valued distribution in a formal parameter $z$:
\begin{equation}
A(z)=\sum_{j\in\mathbb{Z}}  \frac{1}{z^{j+1} } A_{(j)},\quad \text{where } A_{(j)}\in \text{End}(V) ~, 
\end{equation}
and, for all $B \in V$, $A(z)B$ contains only finitely many negative powers of $z$. 

A \textit{vertex algebra} is a vector space $V$ (the \textit{space of states}), 
with a vector $|0\rangle \in V$ (the \textit{vacuum}), 
a map  $Y$ from a given state $A \in V$ to a field $Y(A,z)$ (called  the  \textit{state-field correspondence}), 
and  an endomorphism $\partial: V\rightarrow V$ (the \textit{translation operator}). 
The field $Y(A,z)$ will also be denoted by $A(z)$.

These structures must fulfill a set of axioms. The vacuum should be invariant under translations:  $\partial |0\rangle = 0$. Acting with $\partial$ on a field should be the same as differentiation of the field with respect to the formal parameter~$z$:
\begin{equation} \label{axiom:translinv}
 {[}\partial, Y(A,z)] = \partial_z Y(A,z)~.
\end{equation}
We will use $\partial$ to denote both the endomorphism and $\partial_z$, and it should be clear from the context 
what we mean by $\partial$. 
The field $Y(A,z)$ corresponding to a given state $A$ creates the same state from the vacuum in the limit $z \rightarrow 0$: 
\begin{equation}
Y(A,z) |0\rangle |_{z=0} = A_{(-1)}|0\rangle = A ~.
\end{equation}
The construction easily extends to the case when $V$ is a super vector space. The state-field correspondence $Y$ should respect this grading, $\partial$ should be an even endomorphism, and the vacuum should be even. 

From the endomorphisms $A_{(j)}$ of $Y(A,z)$ (called the \textit{Fourier modes}), we can define the \textit{$\lambda$-bracket}:
\begin{equation} \label{eq:lambdabracketdef}
\LB[\lambda]{A}{B}  =  \sum_{j \geq 0}   \frac{\lambda^j}{j!}  (A_{(j)}B)   ~,
\end{equation}
where $\lambda$ is an even formal parameter.
The $\lambda$-bracket can be viewed as a formal Fourier transformation of $Y(A,z)B$:
\begin{equation} \label{eqlambdabracketdef}
\LB[\lambda]{A}{B} =  \text{Res}_z \; e^{\lambda z} \; Y(A,z)B ~,
\end{equation}
where $\text{Res}_z$ picks the $z^{-1}$-part of the expression. 
The locality axiom of the vertex algebra says that the sum \eqref{eq:lambdabracketdef} is finite for all $A$ and $B$, in other words, all fields in a vertex algebra are mutually local.

The $\lambda$-bracket captures the \textit{operator product expansion} of the corresponding (chiral) fields in a two dimensional quantum field theory. Taking the residue in \eqref{eqlambdabracketdef} picks out the pole in $z$, which can be considered to be a formal $\delta$-function. The parameter $\lambda$ then keeps track of how many derivatives act on the $\delta$-function. In other words, \eqref{eq:lambdabracketdef} is equivalent, in the familiar notation of OPEs, to 
\begin{equation}
A(z) \cdot B(w) \sim \sum_{j \geq 0} \frac{\left( A_{(j)}B \right) (w)}{(z-w)^j}~.
\label{eq:agregado}
\end{equation}

\subsection{$N_K=2$ supersymmetric vertex algebra.} A vertex algebra endowed  with extra supersymmetries can conveniently be described by the formalism of SUSY vertex algebras. By introducing two additional formal parameters, $\theta^1$ and $\theta^2$, that are odd, and promoting the fields~$A(z)$ to \emph{superfields} $A(z,\theta^1,\theta^2)$, we obtain the notion of $N_K=2$ SUSY vertex algebra of \cite{Heluani2007}. In the following, we will often drop the subscript $K$.

We let $Z=(z, \theta^1, \theta^2)$ and consider $N=2$ superfields of the form  
\begin{equation}
A(Z)=\sum_{j\in\mathbb{Z}}  \frac{1}{z^{j+1} } \left(A_{(j|11)} + \theta^1  A_{(j|01)} + \theta^2 A_{(j|10)}+ \theta^1 \theta^2 A_{(j|00)}\right) ~, 
\end{equation}
where $A_{(j|**)} \in \text{End}(V)$
and, as before, for all $B \in V$, $A(Z)B$ contains only finitely many negative powers of $z$. 
The state-field correspondence $Y(A, Z)$ maps a state $A$, to a superfield $A(Z)$. We have two odd endomorphisms: $D_1$ and $D_2$ satisfying $ {[} D_i , D_j ]=\delta_{ij} \partial$ and ${[} D_i , \partial ]=0$.
The vacuum is translation invariant: $D_i |0\rangle = 0$. We require translation invariance,
\begin{equation}\label{axiom:transinv}
 {[}D_i,  Y(A,Z)] = ( \frac{\partial}{\partial \theta^i} - \theta^i \partial_z)  Y(A,Z)~.
\end{equation}

In addition to the even formal parameter $\lambda$, we introduce two odd formal parameters, $\chi_1$ and $\chi_1$, with the relations $[\chi_i,\chi_j]=-2 \delta_{ij}\lambda$ and $[\chi_i,\lambda]=0$. 
We can then define the \textit{N=2 SUSY $\Lambda$-bracket}:
\begin{equation} \label{eq:susylambdabracketdef}
\begin{split} 
\LP{A}{B}  &= \text{res}_Z e^{(z \lambda + \theta^1 \chi_1+ \theta^2 \chi_2)}  Y(A, Z) B \\
&=  \sum_{j \geq 0}   \frac{\lambda^j}{j!} \left(  A_{(j|00)} - \chi_1 A_{(j|10)} + \chi_2 A_{(j|01)} - \chi_1 \chi_2 A_{(j|11)} \right) B~,
\end{split} 
\end{equation}
where $\text{res}_Z$ is the coefficient of $\theta^1 \theta^2 z^{-1}$.
The locality axiom of the SUSY vertex algebra requires that the sum \eqref{eq:susylambdabracketdef} is finite for all $A$ and $B$, \ie, all fields in a SUSY vertex algebra are mutually local.

Let us define the \textit{normal ordered product} $: :$ between two states by
\begin{equation} \label{eq:susynormalorder}
V \otimes V \rightarrow V , \quad A \otimes B \mapsto :AB: \; \equiv A_{(-1 | 11)} B ~.
\end{equation}
In the following, we will often omit the symbol $: :$, and use parenthesis to indicate when the ordering is important. 
Properties of the normal ordering product and the relations between the $\Lambda$-bracket and the normal ordering are given in Appendix \ref{app:rulesLambdabr}. We note however that the normal ordered product  is not associative nor commutative.  The $\Lambda$-bracket and the normally ordered product satisfy a Leibniz-like rule \eqref{QuasiLeibniz} known as the non-commutative Wick formula. In fact, one can define an $N_K=2$ SUSY vertex algebra as a tuple $(V, |0\rangle, :: , [_\Lambda], D^1, D^2, \partial)$ satisfying the axioms of Appendix \ref{app:rulesLambdabr}. 

If one drops the integral terms in the axioms, one arrives to the notion of a \emph{Poisson $N=2$ SUSY vertex algebra} \cite[$\S$ 4.10]{Heluani2007}. In this case, one writes $\{_\Lambda\}$ for the $\Lambda$-bracket, and we note that $V$ with its operation $\cdot$ becomes a unital super-commutative associative algebra since the quantum corrections in \eqref{eq:Quasicommutativity} and \eqref{eq:LambdaBrackeRulesQuasiAssociativity} vanish. Moreover, the Poisson $\Lambda$-bracket $\{_\Lambda\}$ now is distributive with respect to $: :$ (i.e. the Leibniz rule holds) since the quantum correction in \eqref{QuasiLeibniz} vanish. 

Let us consider the situation when one has a family $V_\hbar$ of $N=2$ SUSY vertex algebras parametrized by $\hbar$, that is,  an $N=2$ SUSY vertex algebra over $\mathbb{C}[ [\hbar]]$, such that the fiber at $\hbar=0$ is a Poisson vertex algebra $V_0$ with the operations defined as
\[ : A  B : \; := \lim_{\hbar \rightarrow 0}  :A  B:_\hbar~, \qquad \{A_\Lambda B\} := \lim_{\hbar \rightarrow 0} \frac{1}{\hbar} [A_\Lambda B]_{\hbar} ~. \] 
We then say that the family is a quantization of $V_0$, or that $V_0$ is the \emph{quasiclassical limit} of $V_\hbar$. This happens for example when $V$ is the \emph{universal enveloping SUSY vertex algebra} of a conformal Lie algebra, namely, when $V$ is generated by some fields $A^i$ such that their OPE only involves the fields $A^i$ and their derivatives. In this situation, one may consider the algebra $V_\hbar$ generated by the same $\{A^i\}$ with the $\Lambda$-bracket 
\[ {[A^i}_\Lambda A^j]_\hbar := \hbar {[A^i}_\Lambda A^j]~, \] 
We easily see that the quantum corrections of  \eqref{eq:Quasicommutativity} and \eqref{eq:LambdaBrackeRulesQuasiAssociativity} are of order $\hbar$, and therefore they vanish on $V_0$. We thus obtain a quasiclassical limit of $V_\hbar$.

\subsection{Example: The $\lambda$-brackets of an $N=2$ superconformal vertex algebra.}
The $N=2$ superconformal  vertex algebra is generated by a Virasoro field $L$, two odd fields $G^+$ and $G^-$, an even field $J$, and a central element $c$ (the central charge) \cite{Kac1998}, with 
\begin{align} 
\LB[\lambda]{L}{L} &= (\partial + 2 \lambda ) L + \frac{\lambda^3}{12} c ~, & 
\LB[\lambda]{L}{G^i} &= (\partial + \frac{3}{2} \lambda ) G^i ~,\\
\LB[\lambda]{L}{J} &= (\partial +  \lambda ) J ~, \\
\LB[\lambda]{G^+}{G^-} &= L + (\lambda + \frac{1}{2} \partial)J + \frac{ \lambda^2}{6} c~, &
\LB[\lambda]{G^\pm}{G^\pm} &= 0~, \\
\LB[\lambda]{J}{G^\pm} &= \pm G^\pm~, &
\LB[\lambda]{J}{J} &=  \frac{ \lambda}{3} c~. 
\end{align}

In an $N_K=2$ SUSY vertex algebra, the same algebra is generated by a single field $\G$, with the  $\Lambda$-bracket \cite{Heluani2007}
\begin{equation} \label{eq:N2alg}
\LP{\G}{\G}  = \left( 2 \lambda + 2 \partial + \chi_1 D_1 + \chi_2 D_2 \right) \G + \lambda  \chi_1 \chi_2  \frac{c}{3} ~,
\end{equation}
where the superfield $\G(Z)$ is expanded as
\begin{multline} 
\G(Z) = - i J(z) + i \theta^1 \left(G^+(z) - G^-(z)\right) \\ - \theta^2 \left(G^+(z) + G^-(z) \right) + 2 \theta^1 \theta^2 L(z) ~.
\end{multline}

\subsection{Graded SUSY vertex algebras}

In this section, we recall the concepts of gradings by conformal weights and charge in the supersymmetric case. As always, we restrict to the case of $2$ supersymmetries, and we will omit the terms $N_K=2$ below.  

Recall from \cite[Def. 5.6]{Heluani2007} that a SUSY vertex algebra $V$ is called conformal if it admits a vector $\tau \in V$ such that defining $\G(Z) = Y(\tau, Z)$ this field satisfies \eqref{eq:N2alg}, and moreover
\begin{itemize}
\item $\tau_{(0|00)} = 2 \partial$, $\tau_{(0|10)} = - D_1$, $\tau_{(0|01)} = D_2$.
\item The operator $H:= \tfrac{1}{2} \tau_{(1|00)}$ acts diagonally with eigenvalues bounded below and with finite dimensional eigenspaces. 
\end{itemize}

In this case the eigenvalues of $H$ are called the \emph{conformal weights}. Moreover, it follows from \cite[Thm. 4.16 (4)]{Heluani2007} that, $\forall a \in V$, 
\begin{equation}
[H,Y(a,Z)] = \left( z \partial_z + \frac{1}{2} \left( \theta^1 \partial_{\theta^1} + \theta^2 \partial_{\theta^2} \right) \right) Y(a,Z) + Y(H a, Z) ~.
\label{eq:grading1}
\end{equation}

A SUSY vertex algebra will be called graded if there exists a diagonal operator $H$ satisfying \eqref{eq:grading1}. If $H a = \Delta a$ for $\Delta \in \mathbb{C}$ we say that $a$ has conformal weight, or dimension, $\Delta$. It is easy to see that in this case:
\begin{equation}
\Delta(\partial a) = \Delta(a) + 1, \qquad \Delta(D_i a) = \Delta(a) + \frac{1}{2},  \qquad \Delta(:ab:) = \Delta(a) + \Delta(b), 
\label{eq:grading2}
\end{equation}
and if we let $\Delta(\lambda) = 1$ and $\Delta(\chi^i) = 1/2$ then all the terms of the $\Lambda$ bracket $[a_\Lambda b]$ have conformal weight $\Delta(a) + \Delta(b)$, so that the OPE (or the $\Lambda$ bracket) becomes a graded operation of degree zero. This is a special property of the $N=2$ case, in general the OPE is of degree $N/2 -1$. 

We want to construct a supersymmetric theory where the scalar fields consist of functions on the target manifold. If we want these fields to have dimension zero, then it is clear that their OPE will vanish unless our theory is $N=2$ supersymmetric. In this case, the $\Lambda$-bracket has to be another field of conformal weight zero. In particular, the $\Lambda$-bracket of functions is an operation on functions. 

In fact, the following is a simple exercise in SUSY vertex algebras:

\begin{thm}
Let $V$ be a graded $N_K=2$ SUSY vertex algebra such that the conformal weights are bounded by $0$. Let $V_0$ be the space of conformal weight $0$ vectors, then $V_0$ is naturally a Poisson algebra, with multiplication being the normally ordered product, and the Poisson bracket being the $\Lambda$-bracket.
\label{prop:1}
\end{thm}

Immediately we see that if we want the dimension zero sector of our theory to consists of functions on the target manifold $M$ then $M$ has to be a Poisson manifold. This is the content of the next section. 

The theorem above can be generalized as in the non-SUSY case. Indeed, given a SUSY vertex algebra $V$, it is easy to see that 
\begin{equation}
P(V) := \frac{V}{:V D_1 V: + :V D_2 V:},
\label{eq:poisson2}
\end{equation}
is naturally a Poisson algebra, the associative commutative product is induced from the normally ordered product and the Poisson bracket is induced from the $(0|00)$-th product. If $V$ is graded, then $P\left( V \right)$ inherits the grading, and therefore the zero-th weight space is a Poisson subalgebra. 
\section{Sheaf of $N=2$ VA from a Poisson structure}
\label{s-Def-N2sheaf}

In this section, we construct a sheaf of SUSY vertex algebras on any Poisson manifold $(M, \Pi)$. The heuristic is 
simple; we first attach a local model to an affine space and then we need to prescribe how these local fields change under the allowed local automorphisms (depending on whether we work in the algebraic, real-analytic or smooth setting). In \cite{Malikov1999}, the authors attach to $\mathbb{R}^n$ (in the smooth setting) and coordinates $\{x^\nu\}$ a free $\beta\gamma$-$bc$-system. That is a vertex algebra generated by $2n$ fermionic fields $\{b_\nu, c^\nu\}$, and $2n$ bosonic fields $\{\gamma^\nu, \beta_\nu\}$. What the authors noticed is that under changes of coordinates, the fields $\gamma^\nu$ transform as the coordinates $\{x^\nu\}$ do, the fields $b_\nu$ (respectively $c^\nu$) transform as the vector fields $\partial/\partial_{x^\nu}$ do (respectively the differential forms $dx^\nu$). The fields $\beta_\nu$, however, do not transform as tensorial objects, but in a rather complicated way. In fact, one may think of the generating fields $\gamma^\mu$, $\beta_\mu$, $c^\mu$ and $b_\mu$ as coordinates on the graded supermanifold $\tilde{M}:=T^*[2]T[1]M$ and CDR may be thought of as a formal quantization of loops into this manifold.  

It was noticed in \cite{Ben-Zvi2008} that if we instead of looking at $4n$-fields as generators, we study $2n$-superfields as generators, these objects transform as tensors.  This corresponds to trading supersymmetry in the target by supersymmetry in the worldsheet, namely, instead of loops into $\tilde{M}$ as above, we are looking at $N=1$ superloops (maps from $S^{1|1}$) into the supermanifold $M':=T^*[1]M$. In terms of the previous generators (in the non-SUSY case), the superfields are given by
\[ \phi^\nu= \gamma^\nu + \theta c^\nu, \qquad S_\nu = b_\nu + \theta \beta_\nu~,\]
where the superfields $\phi^\nu$ are even and transform as the coordinates $\{x^\nu\}$ do, while the superfields $S_\nu$ are odd and transform as the vector fields $\partial/\partial_{x^\nu}$ do. 

In this article we exploit further this mechanism by which we trade the complexity of each generator (they are superfields with more components), by simplicity of the transformation formula under changes of coordinates. For this we will look at $N=2$ superloops into $M$. Locally, to $\mathbb{R}^n$ we will attach a SUSY vertex algebra generated by $n$ superfields ($N=2$) $\Phi^\nu$ (which in components account for the $4n$ generators in the classical sense) such that they transform as coordinates do. It follows from Theorem \ref{prop:1} that the OPE of these fields has to be of the form:

%
\begin{equation} \label{N2OPEdefLambdaBracketPois}
\LB{ \Phi^\mu }{ \Phi^\nu} =  \Pi(\Phi)^{\mu \nu} ~,
\end{equation}
where $\Pi^{\mu \nu}$ are the components of a Poisson bivector.  
In fact, we have the following

\begin{thm}
Let $M$ be a Poisson manifold and let $\mathcal{O}$ be its sheaf of smooth functions. There exists a sheaf $\mathcal{V}$ of SUSY vertex algebras on $M$ generated by $\mathcal{O}$, together with an embedding $\iota: \mathcal{O} \rightarrow \mathcal{V}$, such that 
\begin{equation}
\iota (fg) = :\iota(f) \iota(g):, \qquad \iota \{f, g\} = [\iota (f)_\Lambda \iota(g)], 
\label{eq:thm2.1}
\end{equation}
for all local sections $f,g$ of $\mathcal{O}$. This sheaf satisfies a universal property such that for any other sheaf $\mathcal{V}'$ satisfying \eqref{eq:thm2.1}, then there exists a unique surjective morphism $j: \mathcal{V} \rightarrow  \mathcal{V}'$.
\label{thm:2}
\end{thm}
\def\cO{\mathcal{O}}
\begin{proof}
The proof of this statement is straightforward just as in the construction of the chiral de Rham complex \cite{Malikov1999} (see also \cite[Prop.\ 4.6]{Heluani2008a}). Since the construction in the $N=2$ supersymmetric case is simpler than in the non-SUSY case of \cite{Malikov1999} and the $N=1$ case of \cite{Heluani2008a} we sketch here the proof. Locally, one can proceed as follows.  For a Poisson algebra $\cO$ we consider the free $\mathcal{H}$-module generated by $\cO$ (see Appendix~\ref{app:rulesLambdabr} for notation). This module has a structure of SUSY Lie conformal algebra with the operation 
\begin{equation}
[f_\Lambda g] := \{f,g\}~,
\label{eq:thm2.2}
\end{equation}
extended by Sesquilinearity. We can consider its universal enveloping SUSY vertex algebra $V'$ \cite{Heluani2007}. We now consider its quotient $\mathcal{V}$ by the ideal generated by the relations
\begin{equation}
fg = :fg:~, \quad D_i\left( fg \right) := :(D_i f) g: + :f D_i g:~, \quad 1_\mathcal{O} = |0\rangle ~, 
\label{eq:thm2.3}
\end{equation}
$\forall f,g \in \mathcal{O}, \: i=1,2$. Since the operations are defined locally, this ideal is compatible with localization and in fact we obtain a sheaf locally described by this quotient $\mathcal{V}$. 

Notice that $V'$ is naturally graded (declaring $\cO$ to be of degree zero). Since the ideal \eqref{eq:thm2.3} is homogeneous it follows that $\mathcal{V}$ is also graded. In fact, we see that locally $\cO$ is just the degree zero part of $\mathcal{V}$. 
\end{proof}

\begin{rem}
There is a subtlety when we say that this sheaf is generated by $n$-superfields satisfying \eqref{N2OPEdefLambdaBracketPois}. If we are in the algebraic setting and the bivector $\Pi$ is algebraic then we can use arguments of formal geometry to make sense of the RHS of \eqref{N2OPEdefLambdaBracketPois}. In the smooth setting we may construct the sheaf as in the proof of the Theorem, or argue as in \cite{linshaw}.
\label{rem:1}
\end{rem}
This sheaf of  $N_K=2$ SUSY vertex algebras can also be viewed as a sheaf of vertex algebras, generated by the components of $\Phi$. Naming the components of $\Phi$ as
\begin{equation} \label{eq:Phiexpansion}
\Phi^\mu =  \gamma^\mu + \theta_1 c^\mu + \theta_2  d^\mu +  \theta_1  \theta_2 \delta^\mu ~,
\end{equation}
the bracket \eqref{N2OPEdefLambdaBracketPois} is equivalent to the $\lambda$-brackets
\begin{align}
\LB[\lambda]{ \gamma^\mu} { \delta^\nu}&= \Pi^{\mu \nu}~, &
\LB[\lambda]{ c^\mu} { d^\nu}&= \Pi^{\mu \nu}~, \\
\LB[\lambda]{ c^\mu} { \delta^\nu}&= \Pi^{\mu \nu}_{,\tau} c^\tau~, &
\LB[\lambda]{ d^\mu} { \delta^\nu}&= \Pi^{\mu \nu}_{,\tau} d^\tau~, \\
\LB[\lambda]{ \delta ^\mu} { \delta^\nu}&= \Pi^{\mu \nu}_{,\tau \rho} \frac{1}{2}( d^\tau c^\rho - c^\tau d^\rho)~, &
\end{align}
where $\Pi$ is evaluated at $\gamma$ and the rest of the brackets are zero.
 Note that, for a linear Poisson-structure the $\delta$'s  commute. Here $\gamma$ is even, and transforms as a coordinate. The odd fields $c$ and $d$ transforms as vectors, end the even field $\delta$ transforms in an in-homogenous way.

Alternatively, we can generate the sheaf by $N=1$ superfields. Expand $\Phi$ as $\Phi^\mu =  \phi^\mu(z, \theta_1) - \theta_2 S^\mu(z, \theta_1)$. We then have
\begin{align}
\LB{ \phi^\mu}{ S^\nu}_{N_K=1} &= \Pi^{\mu \nu}~, & 
\LB{ S^\mu}{ S^\nu}_{N_K=1} &= \Pi^{\mu \nu}_{~,\tau} S^\tau~.\label{Poisson-Courant-VA}
\end{align}
This shows that the Poisson calculus, in the sense of \cite{vaisman1994lectures}, can be mapped to the $N_K=1$ vertex algebra corresponding to \eqref{N2OPEdefLambdaBracketPois}. For any Poisson manifold, the cotangent bundle is equipped with the non-trivial 
structure of Lie algebroid. Namely, in local coordinates we have
\begin{align}
\{ dx^\mu, dx^\nu \} = \Pi^{\mu\nu}_{~,\tau}dx^\tau~,~~~~~~~~~\{f(x), dx^\mu \} = \Pi^{\mu\nu}\partial_\nu f~,
\end{align}
 where $f(x) \in C^\infty (M)$ and $dx$ is the local basis for differential forms. Thus, on a Poisson manifold one can construct 
 the Courant algebroid (bi-algebroid $TM \oplus T^*M$ with the above bracket on $TM$ and the trivial bracket on $T^*M$)
  and the corresponding sheaf of $N=1$ SUSY vertex algebras is generated by the relations (\ref{Poisson-Courant-VA}). 

\subsection{Relation to the Chiral de Rham complex}

The $N=2$ sheaf can be  related to the Chiral de Rham complex (the sheaf of $N=1$ SUSY vertex algebras associated
 to the standard Courant algebroid on $TM \oplus T^*M$). It is instructive to expand the superfield $\Phi$ in such way so we make contact with previous \cite{Heluani2008a,Ben-Zvi2008,Ekstrand2009c} calculations.

Let $\phi^\mu(z, \theta^1)$ be an even $N=1$ superfield, and $S_\nu(z, \theta^1)$ an odd $N=1$ superfield with the expansions
\begin{equation}
\phi^\mu(z, \theta_1) = \gamma^\mu(z) + \theta^1 c^\mu(z) ~,
\end{equation}
and
\begin{equation}
S_\mu(z, \theta_1) = b_\mu(z) + \theta^1 \beta_\mu(z) ~.
\end{equation}
The field  $\phi^\mu(z, \theta^1)$ transforms as a coordinate, and  $S_\nu(z, \theta^1)$ as a one-form.
Recall that the defining $\Lambda$-bracket of the Chiral de Rham complex is
\begin{equation}\label{eqCDRbrackets}
\LB{ \phi^\mu}{ S_\nu}_{N_K=1} = \delta^{\mu}_{\nu}~, 
\end{equation}
with $\LB{ \phi^\mu}{  \phi^\nu}_{N_K=1}$ and $\LB{ S_\mu}{ S_\nu}_{N_K=1}$ being zero.
Written as $\lambda$-brackets, \eg, with no manifest supersymmetry, this is
\begin{align}
\LB[\lambda]{ \beta_\nu} { \gamma^\mu}&= \delta^{\mu}_{\nu}~, &
\LB[\lambda]{ c^\mu}{ b_\nu}&= \delta^{\mu}_{\nu}~, &
\end{align}
and the rest of the brackets are zero.

From these brackets and fields, we can construct an $N=2$ superfield $\Phi^\mu$, that will fulfill \eqref{N2OPEdefLambdaBracketPois}, by 
\begin{equation} \label{N2expansionN1fields}
\Phi^\mu(z, \theta^1,\theta^2) = \phi^\mu(z, \theta^1) - \theta^2 \Pi^{\mu \nu}( \phi(z, \theta^1) ) S_\nu(z, \theta^1) ~.
\end{equation}
In components, this is
\begin{equation} 
\Phi^\mu =  \gamma^\mu + \theta^1 c^\mu - \theta^2  \Pi^{\mu \nu} b_\nu +  \theta^1  \theta^2(   \Pi^{\mu \nu} \beta_\nu + ( \Pi^{\mu \nu}_{,\tau} c^\tau )  b_\nu  ) ~.
\end{equation}
If the Poisson structure is degenerate, this $\Phi$ may differ from the most general $\Phi$ fulfilling \eqref{N2OPEdefLambdaBracketPois}, and it is only on a symplectic manifold where the sheaf generated by  \eqref{N2OPEdefLambdaBracketPois} is the same as the CDR.

\subsection{Automorphism of the algebra}

The labeling of the two $\theta$'s in the definition of the SUSY vertex algebra is arbitrary, and when we have more then one supersymmetry, we also have an $R$-symmetry. In particular, the bracket \eqref{N2OPEdefLambdaBracketPois} is invariant under the transformations
\begin{align} \label{N2isomorphism}
\theta^1 &\rightarrow   - \theta^2 ~, & \theta^2 &\rightarrow  \theta^1~.
\end{align}
If we also let $D_1 \rightarrow -D_2$ and $D_2 \rightarrow D_1$, then axiom \eqref{axiom:transinv} is still satisfied. 
This automorphism may induce non trivial transformations on the components of the superfields.

\subsection{Quasi-Classical limit} \label{sec:quasi-classical1}
\def\cP{\mathcal{P}}
The sheaf $\mathcal{V}$ constructed above admits a quasi-classical limit $\cP$ as a sheaf of SUSY Poisson vertex algebras. It is generated by $\cO$ just as in \eqref{eq:thm2.1} with the normally ordered product $: :$ replaced by the associative commutative product of the Poisson vertex algebra and its $\Lambda$-bracket $[_\Lambda]$ replaced by the Poisson $\Lambda$-bracket $\{_\Lambda\}$.

\section{$N=2$ algebra on a symplectic manifold}\label{s-symplectic}

In this section, we discuss the case of a symplectic structure. If the Poisson bivector $\Pi$ is invertible, then $M$ is symplectic and 
 we will use a different notation for this case: $\Pi^{\mu\nu} = \omega^{\mu\nu}$. The symplectic structure $\omega_{\mu\nu}$ is a closed non-degenerate 
  two form, such that $\omega^{\mu\nu} \omega_{\nu\rho} = \delta^\mu_\rho$. We can then associate a sheaf of $N=2$ 
  vertex algebras to the manifold, generated by  
\begin{equation} \label{N2defLambdaBracketSympl}
\LB{ \Phi^\mu }{ \Phi^\nu} =  \omega(\Phi)^{\mu \nu} ~.
\end{equation}
  The symplectic case is interesting since we have a canonical two form $\omega_{\mu\nu}$. 
 From the $\Phi$'s, we can construct objects that transforms as vectors, $D_i \Phi^\mu$, or $\partial \Phi^\mu$.  To construct target space diffeomorphism invariant operators, currents, out of these objects, we need tensors with covariant indicies that we can contract with, \eg, forms. The most apparent example to study is the case of a symplectic manifold.

As noted above, this sheaf is essentially the same as the Chiral de Rham complex. If we expand $\Phi$ as in  \eqref{N2expansionN1fields}, we can use $\omega$ to project out $S_\nu$. The brackets  \eqref{eqCDRbrackets} and  \eqref{N2defLambdaBracketSympl} are then equivalent.

The automorphism \eqref{N2isomorphism} induces an automorphism on the components of $\Phi$, given by
\begin{align}
\gamma^\mu &\rightarrow \gamma^\mu ~,&
\beta_\mu &\rightarrow  \beta_\mu +  ( \omega_{\tau \nu, \mu} \omega^{\nu \sigma})(c^\tau b_\sigma)  + \omega_{\mu \sigma, \nu} \partial \omega^{\nu \sigma}~,\\
c^\mu &\rightarrow - \omega^{\mu \nu} b_\nu ~,&
b_\mu &\rightarrow  \omega_{\mu \nu} c^\nu ~.
\end{align}
This automorphism of the $\beta\gamma-bc$-system was discovered, in the case of a Calabi-Yau target manifold, in \cite[Theorem 6.4]{Heluani2008a}.
\def\cV{\mathcal{V}}

\subsection{$N=2$ superconformal algebra}

On the symplectic manifold, the sheaf carries the structure of an $N=2$ superconformal algebra. We can construct a generator $\G_\omega$ by 
\begin{equation} \label{Gdiagonal}
\G_\omega = \frac{1}{2} \omega_{\mu \nu} \left( D_1 \Phi^\mu D_1 \Phi^\nu + D_2 \Phi^\mu D_2 \Phi^\nu \right)  ~.
\end{equation}
There are no order ambiguities in this expression. The operator $\G_\omega$ is a well defined section of the sheaf, and there is no need for any quantum corrections.   The operator fulfill the $N=2$ superconformal algebra
\begin{equation}
\LP{\G_\omega}{\G_\omega}  = \left( 2 \lambda + 2 \partial + \chi_1 D_1 + \chi_2 D_2 \right) \G_\omega + \lambda  \chi_1 \chi_2  \frac{c}{3} ~,
\end{equation}
with central charge $c=3 \;\text{dim} M$.
The proof is given in Appendix \ref{proofGdiagN2alg}.

\section{$N=(2,2)$ vertex algebra on a Calabi-Yau manifold}\label{s-CY}

Let us consider a K\"ahler manifold $M$, with Kähler form $\omega$. Consider the $N_K=2$ SUSY vertex algebra generated by
\begin{equation} \label{eq:LambdaDiracbracket}
\LB{\Phi^\alpha}{ \Phi^{\bar \beta} } = \omega^{\alpha \bar \beta} ~.
\end{equation}
Here the fields $\Phi^\alpha$ and  $\Phi^{\bar \beta}$ correspond to holomorphic and anti-holomorphic coordinates. 
%
Let us define an operator $\mathcal{H}_0$ by
\begin{equation}\label{eq:hamiltonianVA}
\mathcal{H}_0= (g_{\alpha \bar \beta} D_2 \Phi^\alpha ) D_1 \Phi^{\bar \beta}-  ( g_{\alpha \bar \beta} D_1 \Phi^ \alpha ) D_2 \Phi^{\bar \beta}  ~.
\end{equation}
As it stands, this operator is not a well-defined section of the sheaf of vertex algebras for a general Kähler manifold. It may need a ''quantum correction'', as we will see soon. At this stage, the operator might seem rather ad-hoc, but we will motivate it by the discussion of sigma model in section \ref{sec:LagtoHam}.


In order to construct a well defined section of the sheaf of vertex algebras, we need to investigate how $\mathcal{H}_0$ transforms under coordinate changes.
Let $\{z^\alpha\}$ be a holomorphic coordinate system, and let $\tilde{z}^\alpha = F^\alpha(z^\beta)$ be an invertible, holomorphic change of coordinates. We have $\tilde{g}_{\delta \bar \varepsilon}  = g_{\alpha \bar \beta}  \Phi^\alpha_{,\delta} \Phi^{\bar \beta}_{,\bar \varepsilon}$ and
\begin{equation}
\tilde{g}_{\delta \bar \varepsilon}   D_2 \tilde{\Phi}^\delta =(g_{\alpha \bar \beta}  \Phi^\alpha_{,\delta} \Phi^{\bar \beta}_{,\bar \varepsilon} )(\tilde{\Phi}^\delta_{,\gamma}D_2 \Phi^ \gamma) =g_{\alpha \bar \beta}  \Phi^{\bar \beta}_{,\bar \varepsilon} D_2 \Phi^ \alpha ~,
\end{equation}
and, using quasi-associativity \eqref{eq:LambdaBrackeRulesQuasiAssociativity},
\begin{equation}
\begin{split}
\left( \tilde{g}_{\delta \bar \varepsilon}   D_2 \tilde{\Phi}^\delta \right ) D_1 \tilde{\Phi}^{\bar \varepsilon} &= \left( g_{\alpha \bar \beta}  \Phi^{\bar \beta}_{,\bar \varepsilon} D_2 \Phi^\alpha \right )  \left( \tilde{\Phi}^{\bar \varepsilon}_{,\bar \rho}D_1 \Phi^{\bar \rho}\right ) \\
&=  \left( \left( g_{\alpha \bar \beta}  \Phi^{\bar \beta}_{,\bar \varepsilon} D_2 \Phi^\alpha \right ) \tilde{\Phi}^{\bar \varepsilon}_{,\bar \rho}\right ) D_1 \Phi^{\bar \rho} \\
&\quad - 
\left( \int_0^{\nabla} d\Lambda \tilde{\Phi}^{\bar \varepsilon}_{,\bar \rho} \right)  \LB{ g_{\alpha \bar \beta}  \Phi^{\bar \beta}_{,\bar \varepsilon}  D_2 \Phi^\alpha }{D_1 \Phi^{\bar \rho}} \\
&=  \left(g_{\alpha \bar \beta}  D_2 \Phi^\alpha \right )  D_1 \Phi^{\bar \beta}  -i    \Bigl(  \partial  \tilde{\Phi}^{\bar \varepsilon}_{,\bar \rho} \Bigr) \Phi^{\bar \rho}_{, \bar \varepsilon} ~.
\end{split}
\end{equation}
Therefore, under the inverse change of coordinates $\tilde{z} \rightarrow z$,  $\mathcal{H}_0$ transforms as
\begin{equation}
\mathcal{H}_0 \rightarrow \mathcal{H}_0 - 2 i      \Bigl(\partial  \tilde{\Phi}^{\bar \alpha}_{,\bar \rho} \Bigr) \Phi^{\bar \rho}_{, \bar \alpha}= \mathcal{H}_0 - 2 i \frac{\partial \det \bar{A}}{\det \bar{A}} ~,
\end{equation}
where $A^\alpha_{\hphantom{\alpha}\beta} = \partial \tilde{z}^\alpha/ \partial z^\beta$ is the Jacobian of the change of coordinates and $\bar{A}$ is its complex conjugate. We see immediately that $\mathcal{H}_0$ will define a global section of our sheaf if $M$ is Calabi-Yau. In that case, this section looks like \eqref{eq:hamiltonianVA} in the holomorphic coordinate system where the holomorphic volume form is constant. 

To find the expression for this section in a general holomorphic coordinate system we must add a quantum correction to  $\mathcal{H}_0$ that cancels the inhomogeneous transformations.
On a Calabi Yau manifold, we can write the volume form as $\Omega \wedge \bar \Omega$, where $\Omega$ is a holomorphic volume form, $\Omega = e^{f(z)} dz^1 \wedge \ldots \wedge d z^{d/2}$. 
Under the change of coordinates $\tilde{z} \rightarrow z$, $f$ transforms as a density: 
\begin{equation}
\tilde f = f + \log \det \Phi^{\alpha}_{,\beta}= f - \log \det{ A }~.
\end{equation}
We can use this to cancel the inhomogenious transformation of  $\mathcal{H}_0$.
Thus, in general holomorphic coordinates of a Calabi-Yau manifold,
\begin{equation}\label{eq:hamiltonianVAwelldef}
\mathcal{H}=\mathcal{H}_0 - 2  i \partial  \bar f
 =  (g_{\alpha \bar \beta} D_2 \phi^\alpha ) D_1 \phi^{\bar \beta}-  ( g_{\alpha \bar \beta} D_1 \phi^ \alpha ) D_2 \phi^{\bar \beta}  - 2  i \partial  \bar f
\end{equation}
is a well defined section.

Let us now define $\G_\pm$ by
\begin{equation} \label{eq:G1}
\G_\pm= \G_\omega \mp \frac{1}{2} \mathcal{H} ~,
\end{equation}
where $\G_\omega$ is the operator constructed in \eqref{Gdiagonal}.
Introducing new odd derivatives, $D_\pm$, that are linear combinations of the derivatives $D_1$ and $D_2$, by 
\begin{equation}\label{eq:Dpm}
D_\pm \equiv \frac{1}{\sqrt{2}} (D_1 \mp i D_2) ~,
\end{equation}
we can write \eqref{eq:G1} in a general holomorphic coordinate system as 
\begin{equation}\label{eq:G}
\G_\pm= (  \omega_{\alpha \bar \beta} D_\pm \Phi^\alpha  )  D_\mp \Phi^{\bar \beta} \pm  i \partial  \bar f ~.
\end{equation}

The following is the main result of \cite{Heluani2008a} now stated in manifest $N=2$ formalism. The proof can by found in Appendix \ref{proofGpGmN2alg}. 
\begin{thm} Let $M$ be a Calabi-Yau manifold and $\G_\pm$ be defined by \eqref{eq:G}. 
The sections $\G_\pm$ generate two commuting $N=2$ superconformal algebras, 
\begin{equation} \label{eq:N2algs}
\begin{split}
\LP{\G_\pm}{\G_\pm} & = \left( 2 \lambda + 2 \partial + \chi_1 D_1 + \chi_2 D_2 \right) \G_\pm + \lambda  \chi_1 \chi_2  \frac{c}{3} ~,\\
\LP{\G_\pm}{\G_\mp} & = 0 ~,
\end{split}
\end{equation}
each with a central charge $c=\frac{3}{2} \text{dim} M$.
\end{thm}

\section{The $N=2$ Hamiltonian  of an $N=(2,2)$ supersymmetric sigma model} \label{sec:LagtoHam}

We now want to relate the above discussion to the Hamiltonian treatment of the supersymmetric sigma model, and thereby motivate  the expression \eqref{eq:hamiltonianVA}.
To do this, we consider the classical supersymmetric sigma model, and we derive a Hamiltonian formulation thereof. A similar treatment of the $N=1$ sigma model was initiated in \cite{Zabzine2006a, Bredthauer:2006hf} and its relation 
  to CDR was suggested in \cite{Ekstrand2009c}.  Here, we suggest the similar relation between the $N=(2,2)$ supersymmetric sigma models with a Calabi-Yau target and the sheaf of $N=2$ supersymmetric vertex algebras
   on the same Calabi-Yau. 
 
  We restrict ourself to the $N=(2,2)$ supersymmetric sigma model with the target manifold $M$ being a K\"ahler manifold, which is not 
   the most general sigma model with this amount of supersymmetry. The action functional
    for a classical $N=(2,2)$ supersymmetric sigma model is given by
\begin{equation} \label{eq:N2Lagr1}
S=\int d \sigma d \tau d \theta_+^1 d \theta_-^1   d \theta_+^2 d \theta_-^2~ K(\Phi, \bar \Phi) ~,
\end{equation}
 where the integral performed over $\Sigma^{2|4}$ with even coordinates $t, \sigma$ and four odd $\theta$ coordinates.
  For the sake of simplicity, we assume that $\Sigma =  \mathbb{R} \times S^1$.  $\Phi$ and $\bar{\Phi}$ are maps
   from $\Sigma^{2|4}$ to $M$ which satisfy some first order differential equation
    (see Appendix \ref{app:hamderivation}). In physics,
 $\Phi=\{\Phi^\alpha\}$ is called a chiral superfield, and $\bar \Phi=\{\Phi^{\bar\alpha}\}$ is an anti-chiral superfield.  
$K$ is the K\"ahler potential,  which is defined only locally, but nevertheless the action functional (\ref{eq:N2Lagr1})
 is well-defined.  Upon integration of the odd $\theta$-coordinates, the functional (\ref{eq:N2Lagr1}) reduces to the more 
  familiar form of the non-linear sigma model and its critical points are the generalizations of harmonic maps from 
   $\Sigma$ to $M$. In Appendix \ref{app:hamderivation}
    we set the notation and present some properties of this $N=(2,2)$ model which 
    are needed for the derivation.  For more on supersymmetric sigma models and their applications, the reader may 
     consult  the book \cite{MR2003030}.

We would like to consider the Hamiltonian formulation of \eqref{eq:N2Lagr1}.
By doing a change of the odd variables, and integrating out two of them, the action \eqref{eq:N2Lagr1} can be written as
\begin{equation} \label{eq:actioninhamiltonianform1}
S= \int d \sigma d \tau d \theta^2 d \theta^1 \; \left( i K_{,\alpha}  \partial_0 \phi^\alpha  - \frac{1}{2} \mathcal{H} \right) ~,
\end{equation}
with 
\begin{equation}\label{eq:hamiltonian1}
\mathcal{H}= g_{\alpha \bar \beta} D_2 \phi^ \alpha D_1 \phi^{\bar \beta}-  g_{\alpha \bar \beta} D_1 \phi^ \alpha D_2 \phi^{\bar \beta}  
\end{equation}
 being the Hamiltonian and $\partial_0$ being the derivative along $\tau$ (time).
Here, $\theta^1$ and $\theta^2$, with corresponding odd derivatives, $D_i=\frac{\partial}{\partial \theta^i} + \theta^i\partial_\sigma$, are the remaining two odd coordinates. Also, $K_{, \alpha \bar \beta} = g_{\alpha \bar \beta}$.
See Appendix \ref{app:hamderivation} for a more detailed derivation.

From \eqref{eq:actioninhamiltonianform1}, we see that the Poisson bracket is given by
\begin{equation} \label{eqPoissonbr}
\{ \phi^\alpha , \phi^{\bar \beta} \} = \omega^{\alpha \bar \beta} ~,
\end{equation}
and that the Hamiltonian density of the sigma model is given by  \eqref{eq:hamiltonian1}.
The bracket \eqref{eqPoissonbr} is the same as the bracket of the Poisson vertex algebra corresponding to the vertex algebra generated by \eqref{eq:LambdaDiracbracket}. The  expression \eqref{eq:hamiltonian1} is the classical version of the operator $ \mathcal{H}$ considered in \eqref{eq:hamiltonianVAwelldef}.  Thus, following the logic presented in 
  \cite{Ekstrand2009c}, we can think of the sheaf of $N=2$ supersymmetric vertex algebras on a Calabi-Yau as
   a formal quantization of the $N=(2,2)$ sigma model defined by the action (\ref{eq:N2Lagr1}).

\section{Summary and discussion}\label{s-summary}

In this 
note, we construct a sheaf of $N=2$ supersymmetric vertex algebras for a Poisson manifold. We also 
 study the properties of this sheaf on symplectic and Calabi-Yau manifolds. We relate the corresponding
  semiclassical limit  to  the $N=(2,2)$ non-linear sigma model. Let us conclude with a few remarks.

\begin{itemize}

\item As mentioned above, given an $N_K=2$ SUSY vertex algebra $V$, the quotient $P(V)$ defined by \eqref{eq:poisson2} is a Poisson algebra. Just as in the non-SUSY case, there exist an analogous construction of the Zhu algebra associated to $V$, this is a one parameter family of associative superalgebras $P_\hbar(V)$ such that the special fiber $\hbar=0$ coincides with $P(V)$ and all other fibers are isomorphic. In general it is not true that this family is flat, or that $P_\hbar(V)$ is a deformation of the Poisson algebra $P_0(V)$. However, given the construction in this article, starting from a Poisson manifold $M$ with its sheaf of Poisson algebras $\cO$, we constructed a sheaf of SUSY vertex algebras $\cV$ and we obtain a one parameter family of associative algebras $\mathcal{P}_\hbar(\mathcal{V})$. We easily see that $\mathcal{P}_0(\mathcal{V}) = \cO$. 

This immediately leads one to question whether this family is indeed a deformation in this particular case, obtaining thus a natural way of quantizing Poisson manifolds. We plan to return to this topic in a future publication. 

\item The most general $N=(2,2)$ non-linear sigma models are related to generalized K\"ahler geometry 
\cite{Lindstrom:2005zr}. Thus, there should be an analogous Hamiltonian treatment of these general models,
 and it should suggest how to define sheaves of $N=2$ Poisson vertex algebras 
  for a wider class of manifolds. However, it may require a bigger set of fields than considered in this article. This problem remains to be studied. 

\end{itemize}

\section*{Acknowledgement}

\noindent  
M.Z. thanks  KITP, Santa Barbara where part of this work was carried out.
The research of M.Z. is supported by VR-grant 621-2008-4273  and 
 was supported in part by DARPA under Grant No.
HR0011-09-1-0015 and by the National Science Foundation under Grant
No. PHY05-51164.

\appendix
\appendixpage

\section{Rules for \texorpdfstring{$\Lambda$}{Lambda}-brackets in  $N_K=2$ SUSY vertex algebras} \label{app:rulesLambdabr}
In this appendix we collect some properties of $\Lambda$-bracket calculus. For further explanations 
 and details, the reader may consult \cite{Heluani2007}.

\begin{itemize}
\item The operators $D_i$, $\partial$ and the parameters $\chi_j$, $\lambda$, where $i,j=1,2$, have the commutator relations  $[\partial,\chi_i]=[D_i,\lambda]=[\partial,\lambda]=0$,   and 
\begin{align}
[D_i , D_j ]&=2 \delta_{ij} \partial ~,& [\chi_i,\chi_j]&=-2 \delta_{ij}\lambda ~,& [D_i,\chi_j]&=2\delta_{ij}\lambda ~.
\end{align}

We will denote by $\mathcal{H}$ the super-algebra with two odd generators $D_1$, $D_2$ and one even generator $\partial = [D_1,D_2]$ commuting with both $D_1$ and $D_2$.

\item Sesquilinearity:
\begin{subequations}
\begin{align}
\LB{\uD_i a}{ b} &=  - \chi_i \LB{a}{b} ,
& \LB{a}{ \uD_i b} &= (-1)^{a} \left( \uD_i + \chi_i \right) \LB{a}{b} ,
\\
\LB{\partial a }{ b} &=  -\lambda \LB{a}{b},
&\LB{a}{\partial  b} &=  \left( \partial + \lambda \right) \LB{a}{b}  .
\end{align}
\end{subequations}

\item Skew-symmetry:
\begin{equation} 
\LB{a}{b} =  -(-1)^{a b } \LB[-\Lambda - \nabla]{b}{ a} ~.
\label{eq:k.skew.1}
\end{equation}
The bracket on the right hand side is computed as
follows: first compute $\LB[\Gamma]{b}{a}$, where $\Gamma =
(\gamma, \eta)$, then replace $\Gamma$ by $(-\lambda - \partial,
-\chi - \uD)$.

\item Jacobi identity:
\begin{equation}
\LB{a}{\LB[\Gamma]{b}{c}} =  \LB[\Gamma + \Lambda]{ \LB{a}{b}  }{ c} +
(-1)^{ab} \LB[\Gamma]{b}{ \LB{a}{c} } ~.
\label{eq:rulejacobi}
\end{equation}
where the first bracket on the right hand side is computed as in \eqref{eq:k.skew.1}.

An $\mathcal{H}$-module with an operation $[_\Lambda]$ satisfying sesquilinearity, skew-symmetry, and the Jacobi identity is called a SUSY Lie conformal algebra. 

\item Quasi-commutativity:
\begin{equation}
ab - (-1)^{ab} ba = \int_{-\nabla}^0 \LB{a}{b} \ud\Lambda ~,
\label{eq:Quasicommutativity}
\end{equation}
where the integral $\int_{-\nabla}^0 d\Lambda$ is defined as $\frac{\partial}{\partial \chi_1} \frac{\partial}{\partial \chi_2}  \int_{-\partial}^0 d\lambda$.

\item Quasi-associativity:
\begin{equation} \label{eq:LambdaBrackeRulesQuasiAssociativity}
(ab)c - a(bc) =   \left( \int_0^{\nabla} d \Lambda a \right)  \LB{b}{c} + (-1)^{a b} \left( \int_0^{\nabla} d\Lambda b \right)  \LB{a}{c} ~.
\end{equation}

\item Quasi-Leibniz (non-commutative Wick formula):
\begin{equation}
\LB{a}{b c }  = \LB{a}{b} c + (-1)^{ab}b
\LB{ a}{c} + \int_0^\Lambda \LB[\Gamma]{ \LB{a}{b} }{c} \ud \Gamma ~.
\label{QuasiLeibniz}
\end{equation}
\end{itemize}

\section{$N=2$ algebra on a symplectic manifold} \label{proofGdiagN2alg}

We want to show that 
\begin{equation} \label{GdiagonalApp}
\G_\omega = \frac{1}{2} \omega_{\mu \nu} \left( D_1 \Phi^\mu D_1 \Phi^\nu + D_2 \Phi^\mu D_2 \Phi^\nu \right)  
\end{equation}
fulfill
\begin{equation} 
\LP{\G_\omega}{\G_\omega}  = \left( 2 \lambda + 2 \partial + \chi_1 D_1 + \chi_2 D_2 \right) \G + \lambda  \chi_1 \chi_2 \; \text{dim} M ~,
\end{equation}
using the bracket
\begin{equation} 
\LB{ \Phi^\mu }{ \Phi^\nu} =  \omega(\Phi)^{\mu \nu} ~,
\end{equation}
where $\omega^{\mu \nu} \omega_{\nu \tau} = \delta^\mu_\tau$.
Note that there are no ambiguities in the order of the normal ordering in \eqref{GdiagonalApp}. Since each term only contains one type of $D$, there can be no $\chi_1\chi_2$-terms when the brackets of the constituents are calculated. Thus, no terms survive the integration in \eqref{eq:LambdaBrackeRulesQuasiAssociativity}.

We are free to choose any coordinates we want. Since we are on a symplectic manifold, we can choose Darboux coordinates, where $\omega$ is constant. This simplifies the calculations considerably.
Let
\begin{equation}
\G_i \equiv \frac{1}{2} \omega_{\mu \nu} D_i \Phi^\mu D_i \Phi^\nu~.
\end{equation}
We first want to calculate $\LB{\G_i}{\G_i}$. We have 
$\LB{D_i \Phi^\mu}{D_i \Phi^\nu} = \lambda \; \omega^{\mu \nu}$, so $\LB{D_i \Phi^\mu}{\G_i} = \lambda \; D_i \Phi^\mu$. Skew-symmetry then gives
\begin{equation}
\LB{\G_i}{D_i \Phi^\mu} = (\lambda + \partial) \; D_i \Phi^\mu ~.
\end{equation}
From this, we see that
\begin{equation}
\LB{\G_i}{\G_i} = (2 \lambda + \partial) \; \G_i ~.
\end{equation}
We now want to calculate $\LB{\G_1}{\G_2}$. We have
$\LB{D_2 \Phi^\mu}{\G_1} = - \chi_2 \chi_1\; D_1 \Phi^\mu$. Using skew-symmetry, we then get
\begin{equation}
\LB{\G_1}{D_2 \Phi^\mu} =\left (\partial +  \chi_2 D_2 + \chi_1 D_1 \right) D_2 \Phi^\mu +  \chi_2 \chi_1\; D_1\Phi^\mu~.
\end{equation}
From this we see that
\begin{multline} \label{eq:G1DpDp}
\LB{\G_1}{D_2 \Phi^\mu D_2 \Phi^\nu} = \left (\partial +  \chi_2 D_2 + \chi_1 D_1 \right) \left ( D_2 \Phi^\mu D_2 \Phi^\nu \right)   \\
+ \chi_2 \chi_1 \;  \left ( D_1 \Phi^\mu D_2 \Phi^\nu +  D_2 \Phi^\mu D_1 \Phi^\nu \right) + \int ~,
\end{multline}
where the integral term is given by
\begin{equation}
\begin{split}
\int_0^\Lambda \LB[\Gamma]{ \left (\partial +  \chi_2 D_2 + \chi_1 D_1 \right) D_2 \Phi^\mu +  \chi_2 \chi_1\; D_1\Phi^\mu }{D_2 \Phi^\nu} d\Gamma = \\
- \int_0^\Lambda \chi_1 \chi_2 \LB[\Gamma]{D_1\Phi^\mu }{D_2 \Phi^\nu} d\Gamma = - \lambda  \chi_1 \chi_2 \omega^{\mu \nu} ~.
\end{split}
\end{equation}
From \eqref{eq:G1DpDp}, it is now easy to see that 
\begin{multline}
\LB{\G_1}{\G_2} =  \left (\partial +  \chi_2 D_2 + \chi_1 D_1 \right) \G_2 +\\ \chi_2 \chi_1   \omega_{\mu \nu}  D_1 \Phi^\mu D_2 \Phi^\nu +  \lambda  \chi_1 \chi_2 \frac{\text{dim } M}{2},
\end{multline}
and, finally,
\begin{equation} 
\begin{split}
\LP{\G_\omega}{\G_\omega} &= \LP{\G_1}{\G_1} + \LP{\G_2}{\G_2} + \LP{\G_1}{\G_2} + \LP{\G_2}{\G_1}  \\
&=   \left( 2 \lambda + 2 \partial + \chi_1 D_1 + \chi_2 D_2 \right) \G + \lambda  \chi_1 \chi_2 \; \text{dim} M ~.
\end{split}
\end{equation}

\section{$N=(2,2)$ algebra on a Calabi-Yau manifold} \label{proofGpGmN2alg}

We want to calculate the algebra generated by $\G_+$ and $\G_-$, defined in \eqref{eq:G}, under the bracket \eqref{eq:LambdaDiracbracket}. We are free to work in any coordinates we want. A convenient choice is to choose the coordinates where the holomorphic volume form is constant. On a Calabi-Yau, we can always choose such coordinates locally. In this coordinates, the quantum correction $\pm i \partial  \bar f(z)$ vanishes. Also note that $\Gamma^\alpha_{\alpha \beta} = 0$ in these coordinates.
To the metric, we have a corresponding Kähler potential $K$.
Let subscripts of $K$ denote derivatives: $K_{\mu_1 \ldots \mu_k} \equiv \partial_{\mu_1} \ldots \partial_{\mu_k} K$, so $K_{\alpha \bar \beta} = g_{\alpha \bar \beta} = i  \omega_{\alpha \bar \beta}$, with $g$ being the metric and $\omega$ the Kähler form of the manifold. 

Let us define $p_\alpha \equiv i K_{\alpha}$, and
\begin{align}
\G^0_\pm &= D_\mp p_\alpha  D_\pm \phi^\alpha ~, & \M =  i K_{\alpha \beta} \Dp \phi^\alpha \Dm \phi^\beta ~.
\end{align}
We then have  
\begin{equation}\label{eq:GeqG0M} 
\G_\pm = \G^0_\pm \pm  \M ~.
\end{equation}
Note that $\M$ vanishes for a flat manifold .
The definition of $p$ implies the brackets
\begin{subequations} \label{eq:LambdaDiracbrackets}
\begin{align}
\LB{ \phi^\alpha }{ p_\beta } &= \delta_\beta^\alpha ~,&
\LB{ \phi^{\bar\alpha} }{ p_\beta } &= i  \omega^{\bar \alpha \alpha} K_{ \alpha \beta} ~,
\end{align}
in addition to 
\begin{equation}
\LB{\phi^\alpha}{ \phi^{\bar \beta} } = \omega^{\alpha \bar \beta} ~.
\end{equation}
\end{subequations}
In light of the derivation of the Hamiltonian density in section \ref{sec:LagtoHam}, $p_\alpha$ can be understood as the conjugate momenta to $\Phi^\alpha$, and the brackets \eqref{eq:LambdaDiracbrackets} is the corresponding Dirac brackets, see \eqref{eq:Diracbrackets}.

Let us define linear combinations of $\chi_1$ and $\chi_2$, to better suit the base \eqref{eq:Dpm}:
\begin{equation}\label{eq:chipm}
\chi_\pm = \frac{1}{\sqrt{2}} (\chi_1 \pm i \chi_2) ~.
\end{equation}
The relations between $D_\pm$ and $\chi_\pm$ are
\begin{subequations}\label{eq:Dpmalgebra}\begin{align}
 [D_\pm , D_\mp ] &= 2 \partial ~, &  [\chi_\pm , \chi_\mp ] &=- 2 \lambda ~, & [D_\pm , \chi_\pm ] &= 2 \lambda  ~, \\
[D_\pm , D_\pm ] &= 0 ~, & [\chi_\pm , \chi_\pm ] &= 0 ~, & [D_\pm , \chi_\mp ] &=0 ~.
\end{align}\end{subequations}
Note that the rules of sesquilinearity give
\begin{align}
\LP{D_\pm a}{ b} &=  - \chi_\mp \LP{a}{b} ~, &
\LP{a}{ D_\pm b} &= (-1)^a \left( D_\pm+ \chi_\mp \right) \LP{a}{b} ~.
\end{align}

We want to prove that  $\G_+$ and $\G_-$ gives two commuting $N=2$ superconformal algebras, i.\ e.\ 
\begin{equation}
\begin{split}
\LP{\G_\pm}{\G_\pm} & = \left( 2 \lambda + 2 \partial + \chi_+ \Dp + \chi_- \Dm \right) \G_\pm + \lambda  \chi_1 \chi_2  \frac{d}{2} ~,\\
\LP{\G_\pm}{\G_\mp} & = 0 ~.
\end{split}
\end{equation}
We first prove that  $\G^0_+$ and $\G^0_-$ fulfill the algebra \eqref{eq:N2alg}.
In terms of the split \eqref{eq:GeqG0M}, we then need to prove that
\begin{subequations}
\begin{align}
 \LP{\G^0_\pm}{\M} + \LP{\M}{\G^0_\pm} \pm  \LP{\M}{\M} & = \left( 2 \lambda + 2 \partial + \chi_+ \Dp + \chi_- \Dm \right) \M \label{eqG0pMMG0pMM} ~,\\
 \LP{\G^0_\mp}{\M}  - \LP{\M}{\G^0_\pm} \mp  \LP{\M}{\M}  & = 0 \label{eqG0mMMG0pMM} ~.
\end{align}
\end{subequations}


\subsection{Algebra of  $\G^0_\pm$.}
The calculation of $\LP{\G^0_\pm}{\G^0_\pm} $ is straightforward. We do the calculation for $\G^0_+$, the calculation for $\G^0_-$ can be deduced by exchanging $+$ and $-$.
We have
\begin{subequations}
\begin{align}
\LP{p_\alpha}{\G^0_+} & =  \chi_- \Dm p_\alpha~, &  \LP{\G^0_+}{p_\alpha} & = (\chi_- + \Dp) \Dm p_\alpha ~,\\
\LP{\phi^\alpha}{\G^0_+} & = \chi_+ \Dp \phi^\alpha ~, &  \LP{\G^0_+}{\phi^\alpha} & = (\chi_+ + \Dm) \Dp \phi^\alpha ~,
\end{align}
\end{subequations}
and
\begin{align} 
(\chi_\pm + D_\mp) (\chi_\mp + D_\pm) &= - \chi_\mp \chi_\pm -  D_\pm  D_\mp + 2 \partial + \chi_\pm D_\pm - \chi_\mp D_\mp  ~,
\end{align}
so, remembering that $(D_\pm)^2 = 0$, 
\begin{equation} \label{eq:G0pG0p}
\begin{split}
&\LP{\G^0_+}{\G^0_+} = \LP{\G^0_+}{ \Dm p_\alpha}\Dp \phi^\alpha +  \Dm p_\alpha  \LP{\G^0_+}{\Dp \phi^\alpha}  + \int\\
&\quad= (( \chi_+ + \Dm )(\LP{\G^0_+}{  p_\alpha}))\Dp \phi^\alpha +  \Dm p_\alpha  (( \chi_- + \Dp )\LP{\G^0_+}{ \phi^\alpha} ) + \int\\
&\quad= -  \chi_- \chi_+ \G^0_+ + (( 2 \partial  + \chi_+ \Dp) \Dm p_\alpha)\Dp \phi^\alpha \\
&\quad \quad - \chi_+ \chi_- \G^0_+  + \Dm p_\alpha  (( 2 \partial  + \chi_- \Dm )\Dp \phi^\alpha)  + \int \\
&\quad= (2 \lambda  + 2 \partial + \chi_+ \Dp + \chi_- \Dm )  \G^0_+  + \int ~.
\end{split}
\end{equation}
The integral term is given by
\begin{multline} \label{eq:G0G0I}
 \int \LP[\Gamma]{   \LP{\G^0_+}{ \Dm p_\alpha}   }{  \Dp \phi^\alpha } d \Gamma 
=   \int \LP[\Gamma]{ (2 \partial  + \chi_+ \Dp- \chi_- \chi_+) \Dm p_\alpha   }{  \Dp \phi^\alpha } d \Gamma \\
= -  \int ( \chi_- \chi_+ + 2 \gamma  ) \eta_+ \eta_- \delta^\alpha_\alpha d \Gamma
= i    \int ( \chi_- \chi_+ + 2 \gamma  ) \eta_1\eta_2 d \Gamma \frac{d}{2}  \\
=- i  \lambda  ( \chi_- \chi_+  + \lambda  )  \frac{d}{2}  =   \lambda  \chi_1 \chi_2  \frac{d}{2} ~.
\end{multline}

To see that  $\G^0_+$ and $\G^0_-$ commutes, we note that
\begin{align} \label{eq:chipDsquarezero}
(\chi_\pm + D_\mp)^2 &=0 ~,
\end{align}
so
\begin{subequations}
\begin{align}  \label{eq:G0pG0m1}
 \LP{\G^0_+}{ \Dp p_\alpha} &= ( \chi_- + \Dp )\LP{\G^0_+}{  p_\alpha}= (\chi_- + \Dp)^2 \Dm p_\alpha =0 ~, \\
 \LP{\G^0_+}{ \Dp \phi^\alpha} &= ( \chi_+ + \Dm )\LP{\G^0_+}{  \phi^\alpha}= (\chi_+ + \Dm)^2 \Dp \phi^\alpha =0 ~.
 \end{align}
\end{subequations}
Thus,
\begin{equation} \label{eq:G0pG0m}
\LP{\G^0_+}{\G^0_-} = \LP{\G^0_+}{ \Dp p_\alpha}\Dm \phi^\alpha +  \Dp p_\alpha  \LP{\G^0_+}{\Dm \phi^\alpha}  =0 ~.
\end{equation}
There is no integral term.

\subsection{Algebra of  $\G^0_\pm$ and $\M$.}

Let us define some shorthand notation, and calculate some brackets we are going to use later. Let
\begin{align}
B^{\alpha \beta} &\equiv \Dp\phi^\alpha \Dm\phi^\beta ~,
\end{align}
so $\M$ can be written
\begin{align}
\M = i K_{\alpha \beta } B^{\alpha \beta} ~.
\end{align}
Let
\begin{align} \label{eqdefEpm}
E_\pm &\equiv \Gamma^\sigma_{\alpha \beta} K_{ \sigma \gamma} D_\pm \phi^ \gamma B^{\alpha \beta} ~.
\end{align}
Also, note that
\begin{align}
 \label{eq:brPhiK}  \LP{K_{\alpha \beta }}{\phi^\gamma} &= i \Gamma^\gamma_{\alpha \beta } ~. 
\end{align}

\subsubsection{The bracket $\LP{\M}{\M}$.} \label{sssectionMM}
We want to calculate $\LP{\M}{\M}$. 
Since both the first and second argument of the bracket is the same expression,  $\M$,  we only need to calculate the poles represented by an odd number of $\lambda$'s and $\chi$'s, and from skew-symmetry we can deduce the full answer.
We have
\begin{equation} \label{eq:MM}
\LP{\M}{\M}=
i \LP{\M}{K_{\alpha \beta}} B^{\alpha \beta} + i K_{\alpha \beta}    \LP{\M}{B^{\alpha \beta} }  + \int ~,
\end{equation}
where $\int$ represents the integral term in the quasi-Lebniz. 

\paragraph{First term of \eqref{eq:MM}.}
We start with the first term in  \eqref{eq:MM}, so we want to calculate $ \LP{\M}{K_{\alpha \beta}}$. Now,
\begin{equation} \label{eq:MKbrack}
\LP{K_{\alpha \beta}}{\M} = i \LP{K_{\alpha \beta}}{K_{\gamma \delta}} B^{\gamma \delta}  +  i K_{\gamma \delta} \LP{K_{\alpha \beta}}{ B^{\gamma \delta} } ~.
\end{equation}
We then need to calculate $ \LP{K_{\alpha \beta}}{ B^{\gamma \delta} }$:
\begin{equation} \label{eq:KBBr}
\begin{split}
\LP{K_{\alpha \beta}}{ B^{\gamma \delta} } &= \LP{K_{\alpha \beta}}{ \Dp\phi^\gamma} \Dm\phi^\delta +  \Dp\phi^\gamma \LP{K_{\alpha \beta}}{\Dm\phi^\delta } + \int \\
&=i (\Dp + \chi_-) ( \Gamma_{\alpha \beta}^\gamma )  \Dm\phi^\delta + i \Dp\phi^\gamma  (\Dm + \chi_+) ( \Gamma_{\alpha \beta}^\delta ) + \int .
\end{split}
\end{equation}
The integral term of \eqref{eq:KBBr} is
\begin{equation}
\begin{split}
\int_0^\Lambda \LP[\Gamma]{     \LP{K_{\alpha \beta}}{ \Dp\phi^\gamma}     }{     \Dm\phi^\delta    } d \Gamma =&\int_0^\Lambda \LP[\Gamma]{   (\Dp + \chi_- )  \LP{K_{\alpha \beta}}{ \phi^\gamma}     }{     \Dm\phi^\delta    } d \Gamma \\
=& \int_0^\Lambda i (-\eta_- \eta_+) \LP[\Gamma]{    \Gamma_{\alpha \beta}^\gamma     }{     \phi^\delta    } d \Gamma \\
=&- \lambda \LP[\Gamma]{    \Gamma_{\alpha \beta}^\gamma     }{     \phi^\delta    } ~.
\end{split}
\end{equation}
From \eqref{eq:MKbrack}, using skew-symmetry, we have 
\begin{multline} \label{eq:MKf}
 \LP{\M}{K_{\alpha \beta}} =  \chi_+ \Gamma^\delta_{\alpha \beta } K_{\gamma \delta} \Dp \phi^\gamma - \chi_- \Gamma^\gamma_{\alpha \beta }K_{\gamma \delta}  \Dm \phi^\delta - \lambda \; i K_{\gamma \delta}  \LP[\Gamma]{    \Gamma_{\alpha \beta}^\gamma     }{     \phi^\delta    } \\
 - \Dp \phi^\gamma \Dm  K_{\gamma \delta}  \Gamma^\delta_{\alpha \beta } - \Dp K_{\gamma \delta}  \Gamma^\gamma_{\alpha \beta }  \Dm \phi^\delta + \ldots ~,
\end{multline}
where the dots represents terms with no poles, or no odd derivatives, or containing only terms where $D_\pm$ hits holomorphic $\phi$.
So, using the notation defined in \eqref{eqdefEpm}, 
 the first term in \eqref{eq:MM} is
\begin{equation}
 \chi_+ i  E_+ -  \chi_- i  E_-  + \lambda \;  K_{\gamma \delta}  \LP[\Gamma]{    \Gamma_{\alpha \beta}^\gamma     }{     \phi^\delta    }  B^{\alpha \beta} + \mathcal{O} (\lambda^0) ~.
 \end{equation}

\paragraph{Second term of \eqref{eq:MM}.}
To calculate the second term of \eqref{eq:MM}, 
we first calculate $ \LP{B^{\alpha \beta}}{\M} $ using \eqref{eq:KBBr}: 
\begin{multline}
 \LP{B^{\alpha \beta}}{\M} = i \LP{B^{\alpha \beta}}{K_{\gamma \delta}} B^{\gamma \delta}  \\
=   \chi_+ \Gamma^\alpha_{\gamma \delta } \Dp \phi^\beta  B^{\gamma \delta} - \chi_- \Gamma^\beta_{\gamma \delta } \Dm \phi^\alpha  B^{\gamma \delta} - \lambda \; i  \LP[\Gamma]{    \Gamma_{\gamma \delta}^\alpha     }{     \phi^\beta    }  B^{\gamma \delta} + \mathcal{O}(\lambda^0) ~.
 \end{multline}
The second term of \eqref{eq:MM} then is
\begin{equation}
\chi_+ i  E_+ -  \chi_- i E_-  + \lambda \;  K_{\gamma \delta}  \LP[\Gamma]{    \Gamma_{\alpha \beta}^\gamma     }{     \phi^\delta    }  B^{\alpha \beta}  + \mathcal{O}(\lambda^0) ~.
 \end{equation}

\paragraph{Integral term of \eqref{eq:MM}.}

There will be an integral term in \eqref{eq:MM}, given by
\begin{equation} \label{eq:MMintterm}
i \int_0^\Lambda \LP[\Gamma]{ \LP{\M}{K_{\alpha \beta}}}{B^{\alpha \beta}} d \Gamma ~.
\end{equation}
Skew-symmetry still guaranties that we only need to calculate the poles represented by an odd number of $\lambda$'s and $\chi$'s. The integral gives at least $\lambda$, and the possible poles then are $\lambda$ and $\lambda \chi_1 \chi_2$. Higher poles are not possible due to dimensional arguments. Let $\M K \equiv \LP{\M}{K_{\alpha \beta}}$. We have
\begin{equation} \label{eq:MKB}
 \LP[\Gamma]{\M K}{B^{\alpha \beta}} =  \LP[\Gamma]{\M K}{\Dp\phi^\alpha } \Dm\phi^\beta
 +\Dp\phi^\alpha  \LP[\Gamma]{\M K}{ \Dm\phi^\beta }  + \int ~.
\end{equation}
The  integral term  can not be relevant here, since this would give at least a $\gamma$, and the integration in \eqref{eq:MMintterm} would give at least $\lambda^2$, but the highest possible power of $\lambda$ is one.
Recall that the only terms surviving the integration is the $\eta_+ \eta_-$-terms.

We first calculate the first term in \eqref{eq:MKB}. 
We have
\begin{equation}
\begin{split}
 \LP[\Gamma]{\M K}{\Dp\phi^\alpha } = (\eta_- + \Dp)  \LP[\Gamma]{\M K}{\phi^\alpha }  ~.
\end{split}
\end{equation}
So, we need the $\eta_+$- and $\eta_+ \eta_-$-part of $\LP[\Gamma]{\M K}{\phi^\alpha }$, which can be found by looking at the corresponding terms of $\LP[\Gamma]{\phi^\alpha}{\M K }$. 
These, in turn, can be found by using \eqref{eq:MKf}, and we get
\begin{equation}
\begin{split}
 \LP[\Gamma]{\phi^\alpha } {   \LP{\M}{K_{\alpha \beta}}    }_{\eta_+, \eta_+ \eta_-} =& \eta_+  \LP[\Gamma]{\phi^\alpha } {   K_{\gamma \delta} }   \Gamma^\delta_{\alpha \beta }  \Dp \phi^\gamma \\
 &= -i \eta_+  \Gamma^\alpha_{\gamma \delta }   \Gamma^\delta_{\alpha \beta }  \Dp \phi^\gamma ~,
\end{split}
\end{equation} 
and
\begin{equation}
 \LP[\Gamma]{\M K}{\Dp\phi^\alpha }_{\eta_+ \eta_-} = - i \eta_- \eta_+  \Gamma^\alpha_{\gamma \delta }   \Gamma^\delta_{\alpha \beta }  \Dp \phi^\gamma ~,
\end{equation}
so the relevant part of the first term of  \eqref{eq:MKB} is
\begin{equation}
 - i \eta_- \eta_+  \Gamma^\alpha_{\gamma \delta }   \Gamma^\delta_{\alpha \beta } B^{\gamma  \beta } ~.
\end{equation}
The second term can be calculated by exchanging $+$ and $-$, and yields the same term. In total, the integral terms is \begin{equation}
 \int_0^\Lambda 2   \eta_- \eta_+  \Gamma^\alpha_{\gamma \delta }   \Gamma^\delta_{\alpha \beta }  B^{\gamma \beta} d\Gamma = -2  i \lambda  \; \Gamma^\alpha_{\gamma \delta }   \Gamma^\delta_{\alpha \beta }  B^{\gamma \beta} ~.
\end{equation}

\paragraph{In total.}
Summing the contributions, and using skew-symmetry, we have 
\begin{equation}
\begin{split}
\LP{\M}{\M} =& A +  \chi_+  2 i  E_+ -  \chi_- 2 i E_- +  \lambda \;  2  Q\\
=& -A +  (\chi_+ + \Dm) 2 i E_+ - (\chi_- + \Dp)  2 i E_-  + (\lambda + \partial)  Q ~,
\end{split}
\end{equation}
where $A$ is the part of the bracket with no $\lambda$'s or $\chi$'s, and
\begin{equation}
Q \equiv ( K_{\gamma \delta}  \LP[\Gamma]{    \Gamma_{\alpha \beta}^\gamma     }{     \phi^\delta    }  - i \Gamma^\gamma_{\alpha \delta }   \Gamma^\delta_{\gamma \beta }  ) B^{\alpha \beta} ~.
\end{equation}
Thus
\begin{equation} \label{eq:MMf}
\LP{\M}{\M} =  (2\chi_+ + \Dm) i E_+ -  (2 \chi_- + \Dp) i E_-  + (2\lambda + \partial)  Q ~.
\end{equation}

\subsubsection{The bracket $\LP{\M}{\G^0_+}$.} \label{sssectionGMMG}
We  want to calculate $\LP{\M}{\G^0_+} + \LP{\G^0_+}{\M}$, and later $\LP{\M}{\G^0_-} - \LP{\G^0_+}{\M}$. We start with $\LP{\M}{\G^0_+}$, and the other terms can then be calculated by using skew-symmetry and by exchanging $+$ and $-$.  Using Leibniz and sesquilinearity, we see that
\begin{equation} \label{eq:MGp1}
\begin{split}
\LP{\M}{\G^0_+} =&   \LP{\M}{ \Dm p_\gamma}  \Dp \phi^\gamma    +    \Dm p_\gamma\LP{\M}{\Dp \phi^\gamma} + \int   \\
=&    (\chi_+ \! + \! \Dm)( \LP{\M}{ p_\gamma}   ) \Dp \phi^\gamma  \\
&+  \Dm p_\gamma  (\chi_- \!+\! \Dp)( \LP{\M}{ \phi^\gamma} )+ \int   ~.
\end{split}
\end{equation}

\paragraph{First part  of \eqref{eq:MGp1}.} 
We first note that
\begin{equation}
\LP{p_\gamma}{\M} = i \LP{p_\gamma}{K_{\alpha \beta }} B^{\alpha \beta} + i K_{\alpha \beta }  \LP{p_\gamma}{ B^{\alpha \beta}}  ~,
\end{equation}
with no integral term, and
\begin{equation}
 \LP{p_\gamma}{ B^{\alpha \beta}}  =  
 \chi_+ \delta_\gamma^\beta \Dp\phi^\alpha 
 - \chi_- \delta_\gamma^\alpha \Dm\phi^\beta ~,
\end{equation}
so
\begin{equation}
  \LP{p_\gamma}{\M} = i \LP{p_{\gamma}  }{K_{\alpha \beta }}  B^{\alpha \beta} + i \chi_+ K_{\gamma \alpha } \Dp\phi^\alpha - i \chi_- K_{\gamma \alpha } \Dm\phi^\alpha ~.
\end{equation}
Using skew-symmetry, we have
\begin{equation} \label{eq:Mp}
\begin{split}
\LP{\M}{p_\gamma} =& - i  \LP{p_{\gamma}  }{K_{\alpha \beta }}  B^{\alpha \beta}  \\
&+ i (\chi_+ + \Dm)( K_{\gamma \alpha } \Dp\phi^\alpha) - i (\chi_- + \Dp)( K_{\gamma \alpha } \Dm\phi^\alpha)  ~.
\end{split}
\end{equation}
From \eqref{eq:Dpmalgebra}, we see that $(\chi_\pm + D_\mp)^2 =0$, and we note that
\begin{align}
(\chi_+ + \Dm)(\chi_- + \Dp) &=   \chi_+ \Dp -\chi_-\Dm - \chi_- \chi_+ +  \Dm \Dp  ~,
\end{align}
so, the first part of \eqref{eq:MGp1} is 
\begin{equation} \label{eq:MGp1part1}
\begin{split}
&- i (\chi_+ + \Dm) ( \LP{p_{\gamma}  }{K_{\alpha \beta }}  B^{\alpha \beta})\Dp \phi^\gamma   \\
& - i ( \chi_+ \Dp -\chi_-\Dm - \chi_- \chi_+ +  \Dm \Dp ) ( K_{\gamma \alpha } \Dm\phi^\alpha)  \Dp \phi^\gamma \\
=& - i (\chi_+ + \Dm) ( \LP{p_{\gamma}  }{K_{\alpha \beta }}  B^{\alpha \beta})\Dp \phi^\gamma  +\chi_+ \Dp \M  \\
&- \chi_- \chi_+ \M  - i \chi_-\Dm  K_{\alpha \beta }  B^{\alpha \beta} - i \Dm \Dp ( K_{\gamma \alpha } \Dm\phi^\alpha)  \Dp \phi^\gamma  ~.
\end{split}
 \end{equation}
 
 \paragraph{Second part  of \eqref{eq:MGp1}.} 
We have
\begin{equation} \label{eq:brphiM}
\LP{\phi^\gamma}{\M} = i \LP{\phi^\gamma}{K_{\alpha \beta }} B^{\alpha \beta} =  \Gamma^\gamma_{\alpha \beta }  B^{\alpha \beta}  ~,
\end{equation}
so, the second part of \eqref{eq:MGp1} is 
\begin{equation} \label{eq:MGp1part2tmp}
-  \Dm p_\gamma (\chi_-  +  \Dp)( \Gamma^\gamma_{\alpha \beta }  B^{\alpha \beta}  ) ~.
\end{equation}
In the coordinates chosen, this is
\begin{equation} \label{eq:MGp1part2}
   (\chi_-  +  \Dp)( \Gamma^\gamma_{\alpha \beta }  B^{\alpha \beta}  ) \Dm p_\gamma - i \partial (\LP{ p_\gamma }{   \Gamma^\gamma_{\alpha \beta} }   B^{\alpha \beta}) ~.
\end{equation}

\paragraph{Integral term  of \eqref{eq:MGp1}.} 
The integral term in  \eqref{eq:MGp1} is given by
\begin{equation}
\int_0^\Lambda \LP[\Gamma]{    \LP{\M}{ \Dm p_\gamma}  }{  \Dp \phi^\gamma} d \Gamma ~.
\end{equation}
We neeed
\begin{equation}
\begin{split}
\LP[\Gamma]{    \LP{\M}{ \Dm p_\gamma}  }{  \Dp \phi^\gamma} &=- (\eta_- + \Dp) \LP[\Gamma]{    \LP{\M}{ \Dm p_\gamma}  }{  \phi^\gamma}  \\
&=- (\eta_- + \Dp) \LP[\Gamma]{  (\Dm + \chi_+)  \LP{\M}{ p_\gamma}  }{  \phi^\gamma} \\
&=(\eta_- + \Dp)(\eta_+  + \chi_+) \LP[\Gamma]{ \LP{\M}{ p_\gamma}  }{  \phi^\gamma} ~.
\end{split}
\end{equation}
Using \eqref{eq:Mp} we get
\begin{equation}
\begin{split}
\LP[\Gamma]{    \phi^\gamma }{    \LP{\M}{  p_\gamma}   } =& - i \LP[\Gamma]{    \phi^\gamma }{   \LP{p_{\gamma}  }{K_{\alpha \beta }}  }  B^{\alpha \beta} \\
&+ i (\Dm + \eta_+ + \chi_+ )  \LP[\Gamma]{    \phi^\gamma }{ K_{\gamma \alpha }}  \Dp\phi^\alpha \\
&- i (\Dp + \eta_- + \chi_- )  \LP[\Gamma]{    \phi^\gamma }{ K_{\gamma \alpha }}  \Dm\phi^\alpha ~.
\end{split}
\end{equation}
We have $ \LP[\Gamma]{    \phi^\gamma }{ K_{\gamma \alpha }} = \Gamma^\gamma_{\gamma \alpha} = 0$, so
\begin{equation}
\LP[\Gamma]{    \LP{\M}{  p_\gamma}   }{    \phi^\gamma } =  i \LP[\Gamma]{    \phi^\gamma }{   \LP{p_{\gamma}  }{K_{\alpha \beta }}  }  B^{\alpha \beta} ~,
\end{equation}
and
\begin{equation}
\LP[\Gamma]{    \LP{\M}{ \Dm p_\gamma}  }{  \Dp \phi^\gamma} = i  \eta_- \eta_+  \LP[\Gamma]{    \phi^\gamma }{   \LP{p_{\gamma}  }{K_{\alpha \beta }}  }  B^{\alpha \beta} ~,
\end{equation}
so the quantum correction is
\begin{equation}
 \lambda \LP[\Gamma]{    \phi^\gamma }{   \LP{p_{\gamma}  }{K_{\alpha \beta }}  }  B^{\alpha \beta} ~.
\end{equation}
Using Jacobi, this is
\begin{equation} \label{eq:integraltermMGp}
 \lambda \LP{  p_{\gamma} }{   \LP[\Gamma]{  \phi^\gamma  }{K_{\alpha \beta }}  }  B^{\alpha \beta}  = - i  \lambda   \LP{  p_{\gamma} }{ \Gamma^\gamma_{\alpha \beta }  }  B^{\alpha \beta}  ~.
\end{equation}

\paragraph{In total.}
Thus,  \eqref{eq:MGp1} is the sum of  \eqref{eq:MGp1part1}, \eqref{eq:MGp1part2} and \eqref{eq:integraltermMGp}:
\begin{multline}
   \chi_+ \Dp \M  - \chi_- \chi_+ \M 
+ \chi_- ( \Gamma^\gamma_{\alpha \beta }  B^{\alpha \beta}   \Dm p_\gamma - i \Dm K_{ \alpha \beta } B^{\alpha \beta}) \\
-i  \chi_+ (   \LP{p_{\gamma}  }{K_{\alpha \beta }}  B^{\alpha \beta} \Dp \phi^\gamma  )  + \Dp( \Gamma^\gamma_{\alpha \beta }  B^{\alpha \beta}  ) \Dm p_\gamma  \\
  -i \Dm \Dp  ( K_{\gamma \alpha } \Dm\phi^\alpha)  \Dp \phi^\gamma
-i  \Dm ( \LP{p_{\gamma}  }{K_{\alpha \beta }}  B^{\alpha \beta})\Dp \phi^\gamma \\
-( \lambda + \partial)( i  \; \LP{ p_\gamma }{   \Gamma^\gamma_{\alpha \beta} }   B^{\alpha \beta}) ~.
 \end{multline}
We have
\begin{align}
i \LP{p_{\gamma}  }{K_{\alpha \beta }}  B^{\alpha \beta}  \Dp \phi^\gamma& = i E_+ ~, \\
\Gamma^\gamma_{\alpha \beta }  B^{\alpha \beta}   \Dm p_\gamma -  i \Dm K_{ \alpha \beta } B^{\alpha \beta} &=  i E_-  ~.
\end{align}
Also,
\begin{equation}
\begin{split}
-i \Dm \Dp  ( K_{\gamma \alpha } \Dm\phi^\alpha)  \Dp \phi^\gamma =&- i \Dp ( \Dm  K_{ \alpha \beta} B^{\alpha \beta} ) \\
&+ i2  \partial K_{ \alpha \beta } B^{\alpha \beta}+ i 2 K_{ \alpha \beta } \Dp \phi^ \alpha \partial \Dm\phi^\beta ~,
\end{split}
\end{equation}
so
\begin{equation} \label{eq:MGp2}
\begin{split}
\LP{\M}{\G^0_+} =&   -  \chi_- \chi_+ \M +   \chi_+ \Dp \M  + i (\chi_- +\Dp) E_- - i ( \chi_++ \Dm) E_+  \\
& + 2 i \partial K_{ \alpha \beta } B^{\alpha \beta}+2  i   K_{ \alpha \beta } \Dp \phi^ \alpha \partial \Dm\phi^\beta \\
&-  \Gamma^\gamma_{\alpha \beta }  B^{\alpha \beta}  \Dp \Dm p_\gamma
+ i  \LP{p_{\gamma}  }{K_{\alpha \beta }}  B^{\alpha \beta} \Dm\Dp \phi^\gamma \\
&- (\lambda + \partial)( i  \; \LP{ p_\gamma }{   \Gamma^\gamma_{\alpha \beta} }   B^{\alpha \beta}) ~.
\end{split}
\end{equation}

\paragraph{$\LP{\G^0_+}{\M}$, and taking the sum.}
Using skew-symmetry, from \eqref{eq:MGp2}, we calculate $\LP{\G^0_+}{\M}$:
\begin{equation} \label{eq:GpM}
\begin{split}
\LP{\G^0_+}{\M} =&   \chi_- \chi_+ \M +  2 \lambda  \M  + 2 \partial \M  +  \chi_- \Dm \M  + i \chi_-  E_- - i  \chi_+ E_+  \\
& -2  i  \partial K_{ \alpha \beta } B^{\alpha \beta}-2  i  K_{ \alpha \beta } \Dp \phi^ \alpha \partial \Dm\phi^\beta \\
&+ \Gamma^\gamma_{\alpha \beta }  B^{\alpha \beta}  \Dp \Dm p_\gamma
- i \LP{p_{\gamma}  }{K_{\alpha \beta }}  B^{\alpha \beta} \Dm\Dp \phi^\gamma\\
&- \lambda  i  \; \LP{ p_\gamma }{   \Gamma^\gamma_{\alpha \beta} }   B^{\alpha \beta}   ~.
\end{split}
\end{equation}
Taking the sum of \eqref{eq:MGp2} and \eqref{eq:GpM}, we have
\begin{equation} \label{eqG0pMMG0pf}
\begin{split}
\LP{\G^0_+}{\M} + \LP{\M}{\G^0_+}  =&  (2 \lambda  + 2 \partial  + \chi_- \Dm  +  \chi_+ \Dp) \M  \\
&+ i (2\chi_- + \Dp)  E_- - i (2  \chi_+ + \Dm) E_+ \\
&- (2 \lambda + \partial) ( i  \; \LP{ p_\gamma }{   \Gamma^\gamma_{\alpha \beta} }   B^{\alpha \beta} )  ~.
\end{split}
\end{equation}

\subsubsection{Summing the results from \ref{sssectionMM} and \ref{sssectionGMMG}}
We want to show that  \eqref{eqG0pMMG0pMM} is fulfilled.
From \eqref{eq:MMf} and \eqref{eqG0pMMG0pf}, we get
\begin{multline}
 \LP{\G^0_+}{\M} + \LP{\M}{\G^0_+} +  \LP{\M}{\M} = \left( 2 \lambda + 2 \partial + \chi_+ \Dp + \chi_- \Dm \right) \M \\
 - (2\lambda + \partial)\left (\left( -K_{\gamma \delta}  \LP[\Gamma]{    \Gamma_{\alpha \beta}^\gamma     }{     \phi^\delta    }  + i \Gamma^\gamma_{\alpha \delta }   \Gamma^\delta_{\gamma \beta }  +  i   \LP{ p_\gamma }{   \Gamma^\gamma_{\alpha \beta} } \right) B^{\alpha \beta} \right) ~.
\end{multline}
The parenthesis of the last line is
\begin{equation} \label{eq:zeroRiemanntensorholomorp}
- K_{\gamma \delta}  \LP[\Gamma]{    \Gamma_{\alpha \beta}^\gamma     }{     \phi^\delta    }  + i \Gamma^\gamma_{\alpha \delta }   \Gamma^\delta_{\gamma \beta }  +  i   \LP{ p_\gamma }{   \Gamma^\gamma_{\alpha \beta} } =  i \Gamma^\gamma_{\alpha \delta }   \Gamma^\delta_{\gamma \beta } - i \partial_\gamma \Gamma^\gamma_{\alpha \beta}  = 0~.
\end{equation}
This is zero in the coordinates chosen. So,  \eqref{eqG0pMMG0pMM} is fulfilled. The corresponding equation for the $-$-sector comes by exchanging $+$ and $-$. We have thus shown that
\begin{equation}
\LP{\G_\pm}{\G_\pm}  = \left( 2 \lambda + 2 \partial + \chi_+ \Dp + \chi_- \Dm \right) \G_\pm + \lambda  \chi_1 \chi_2  \frac{d}{2} ~.
\end{equation}
We now want to show that $\G_+$ and $\G_-$ commute.

\subsubsection{ $\LP{\G^0_-}{\M} - \LP{\M}{\G^0_+} $   }
We  want to calculate  $\LP{\G^0_-}{\M} - \LP{\M}{\G^0_+} $.
From \eqref{eq:GpM} we see that $\LP{\G^0_-}{\M}$ is
\begin{equation} \label{eq:GmM}
\begin{split}
\LP{\G^0_-}{\M} =&   \chi_+ \chi_- \M +  2 \lambda  \M  + 2 \partial \M  +  \chi_+ \Dp \M  + i \chi_+  E_+ - i  \chi_- E_-  \\
& -2  i \partial K_{ \alpha \beta } B^{\alpha \beta}  +2  i  K_{ \alpha \beta } \Dm \phi^ \alpha \partial \Dp\phi^\beta \\
&+ \Gamma^\gamma_{\alpha \beta }  B^{\alpha \beta}  \Dm \Dp p_\gamma
- i \LP{p_{\gamma}  }{K_{\alpha \beta }}  B^{\alpha \beta} \Dp\Dm \phi^\gamma \\
&+ \lambda  i  \; \LP{ p_\gamma }{   \Gamma^\gamma_{\alpha \beta} }   B^{\alpha \beta}   ~.
\end{split}
\end{equation}
Note that when we exchange $+$ and $-$, or equivalently $1$ and $2$ in the numbering of the supersymmetries, we keep the integration order in the integrals fixed. This yields an extra minus sign in  the quantum term above.

The difference between \eqref{eq:GmM} and \eqref{eq:MGp2} is
\begin{equation} \label{eq:G0mM2}
\begin{split}
&\LP{\G^0_-}{\M} - \LP{\M}{\G^0_+} =   (\chi_+ \chi_-  + \chi_- \chi_+ + 2 \lambda + 2 \partial)  \M \\
&\quad+  i (2 \chi_+ +\Dm) E_+ -  i (2 \chi_- +\Dp) E_-  \\
& \quad -4 i \partial K_{ \alpha \beta } B^{\alpha \beta}-  2 i  K_{ \alpha \beta } \Dp \phi^ \alpha \partial \Dm\phi^\beta - 2 i  K_{ \alpha \beta } \partial \Dp \phi^ \alpha \Dm\phi^\beta \\
& \quad+ \Gamma^\gamma_{\alpha \beta }  B^{\alpha \beta}  (\Dm \Dp + \Dp \Dm) p_\gamma \\
& \quad- i \LP{p_{\gamma}  }{K_{\alpha \beta }}  B^{\alpha \beta}  (\Dm \Dp + \Dp \Dm) \phi^\gamma 
 +i (2 \lambda + \partial)  \;  ( \LP{ p_\gamma }{   \Gamma^\gamma_{\alpha \beta} }   B^{\alpha \beta} )  \\
&=  +  i (2 \chi_+ +\Dm) E_+ -  i (2 \chi_- +\Dp) E_-  \\
& \quad + 2 \partial  \M        
-4 i \partial K_{ \alpha \beta } B^{\alpha \beta}
- 2 i  K_{ \alpha \beta } \partial B^{\alpha \beta} \\
& \quad+ 2 \Gamma^\gamma_{\alpha \beta }  B^{\alpha \beta} \partial p_\gamma
- 2 i  \LP{p_{\gamma}  }{K_{\alpha \beta }}  B^{\alpha \beta}  \partial \phi^\gamma \\
& \quad +i  (2 \lambda + \partial)   \;  ( \LP{ p_\gamma }{   \Gamma^\gamma_{\alpha \beta} }   B^{\alpha \beta} )  ~.
\end{split}
\end{equation}
The third line of \eqref{eq:G0mM2} can be simplified, noting
\begin{equation}
\begin{split}
\Gamma^\gamma_{\alpha \beta }  B^{\alpha \beta} \partial p_\gamma
-  i  \LP{p_{\gamma}  }{K_{\alpha \beta }}  B^{\alpha \beta}  \partial \phi^\gamma &= \\
  i  \Gamma^\gamma_{\alpha \beta }  B^{\alpha \beta}K_{\gamma \sigma} \partial \phi^\sigma +i  \Gamma^\gamma_{\alpha \beta }  B^{\alpha \beta}K_{\gamma\bar \sigma} \partial \phi^{\bar\sigma} \\
-i  \Gamma^\sigma_{\alpha \beta } K_{ \sigma \gamma} B^{\alpha \beta}  \partial \phi^\gamma + i K_{\alpha \beta \gamma} B^{\alpha \beta} \partial \phi^\gamma &=\\
i K_{\alpha \beta \bar\gamma} B^{\alpha \beta} \partial \phi^{\bar\gamma} + i K_{\alpha \beta \gamma} B^{\alpha \beta} \partial \phi^\gamma &=
i \partial K_{\alpha \beta } B^{\alpha \beta} ~.
\end{split}
\end{equation}
So, finally,
\begin{equation} \label{eq:G0mMf}
\begin{split}
\LP{\G^0_-}{\M} - \LP{\M}{\G^0_+} =&  i (2 \chi_+ +\Dm) E_+ -  i (2 \chi_- +\Dp) E_-   \\
&+i (2 \lambda +  \partial )   \;  ( \LP{ p_\gamma }{   \Gamma^\gamma_{\alpha \beta} }   B^{\alpha \beta} )  ~,
\end{split}
\end{equation}
and, using \eqref{eq:MMf} and  \eqref{eq:zeroRiemanntensorholomorp}, 
\begin{multline} 
 \LP{\G^0_-}{\M}  - \LP{\M}{\G^0_+} -  \LP{\M}{\M}  = \\
+ (2\lambda + \partial)\left (\left( -K_{\gamma \delta}  \LP[\Gamma]{    \Gamma_{\alpha \beta}^\gamma     }{     \phi^\delta    }  + i \Gamma^\gamma_{\alpha \delta }   \Gamma^\delta_{\gamma \beta }  +  i  \; \LP{ p_\gamma }{   \Gamma^\gamma_{\alpha \beta} } \right) B^{\alpha \beta} \right) = 0 ~.
\end{multline}
Thus
\begin{equation}
\LP{\G_+}{\G_-}=0 ~.
\end{equation}

\section{Derivation of the Hamiltonian of the  $N=(2,2)$ supersymmetric sigma model} \label{app:hamderivation}

The action for an $N=(2,2)$ supersymmetric sigma model with a K\"ahler target manifold is given by
\begin{equation} \label{eq:N2Lagr}
S=\int d \sigma d \tau d \theta_+^1 d \theta_-^1   d \theta_+^2 d \theta_-^2 K(\Phi, \bar \Phi) ~,
\end{equation}
where $K$ is the K\"ahler potential, and $\Phi=\{\Phi^\alpha\}$ is a chiral superfield, and $\bar \Phi=\{\Phi^{\bar\alpha}\}$ is an anti-chiral superfield.  We use indicies $\mu, \nu, \ldots$ to denote real coordinates, and $\alpha, \beta, \ldots$ to denote complex coordinates.

We have two copies of the $N=(1,1)$ algebra\footnote{Here we misuse the spinor notation. For example, the partial derivative $\partial_+$ should be understood as $\partial_{++}$ in spinor indices. Since we are after the Hamiltonian 
 treatment, the Lorentz covariance is not the issue.}:
\begin{align}
(D_\pm^i)^2 &= i \partial_\pm ~,  & \{ \Dp^i, \Dm^j \} &= 0 ~,& \{ D_\pm^1, D_\pm^2 \} &= 0 ~, & i,j&=1,2 ~,
\end{align}
where
\begin{align}
D_\pm^i &= \frac{\partial}{\partial \theta_\pm^i} + i \theta_\pm^i \partial_\pm , & \partial_\pm = \partial_0 \pm \partial_1 ~.
\end{align}
For chiral and anti-chiral superfields,  the two supersymmetries are related:
\begin{align} \label{eq:chiralrelations}
D_\pm^1 \Phi^{\alpha} &=  i D_\pm^2 \Phi^{\alpha} ~, & D_\pm^1 \Phi^{\bar \alpha} &= - i D_\pm^2 \Phi^{\bar \alpha} ~.
\end{align}
 In physics literature $D_\pm^i$ are typically combined into complex operators and also it is more customary to 
  use the the complex $\theta$'s (for example, see the conventions in \cite{Lindstrom:2005zr}). 

We now want to go to Hamiltonian formalism. We are going to integrate out two odd coordinates, and write \eqref{eq:N2Lagr} in a first order form.

Let us define a new set of $\theta$'s:

\begin{equation}
\begin{bmatrix}
\theta_0^1\\
\theta_0^2\\
\theta_1^1\\
\theta_1^2
\end{bmatrix}
=
\frac{1}{\sqrt{2}}
\begin{bmatrix} 
\hphantom{-}1 &	\hphantom{-}1 &-i&	\hphantom{-}i \\
-i &	\hphantom{-} i &	\hphantom{-}1 & 	\hphantom{-}1 \\
\hphantom{-}1 &	-1 &	\hphantom{-}i &	-i \\
\hphantom{-}i &	-i &	\hphantom{-}1 &	-1
\end{bmatrix}
\begin{bmatrix}
\theta_+^1\\
\theta_-^1\\
\theta_+^2\\
\theta_-^2
\end{bmatrix} ~.
\end{equation}
The action then is
$S= - \frac{1}{2} \int d \sigma d \tau d \theta_1^1 d \theta_1^2   d \theta_0^1 d \theta_0^2 K 
$.
We now want to integrate out $\theta_0^1$ and $\theta_0^2$.
Introduce new differential operators:
\begin{align} 
    D_0^1 &= \tfrac{1}{\sqrt{2}}(\Dp^1 +  \Dm^1) ~, 
& D_0^2 &= \tfrac{1}{\sqrt{2}}(\Dp^2 +  \Dm^2)~, 
& \\
D_1^1 &=\tfrac{1}{\sqrt{2}}(\Dp^1 + i \Dm^2)~, 
& D_1^2 &=\tfrac{1}{\sqrt{2}}(\Dp^2 +i  \Dm^1) ~.
\end{align}
We have
\begin{align}
D_0^1 &=  \frac{\partial}{\partial \theta_0^1} + i \theta_0^1 \partial_0 +  \frac{1}{2} (i \theta_1^1   - \theta_1^2  ) \partial_0 + \frac{1}{2} ( i \theta_1^1   + \theta_1^2  ) \partial_1 ~,\\
D_0^2 &=  \frac{\partial}{\partial \theta_0^2} +i  \theta_0^2 \partial_0 +  \frac{1}{2} (i  \theta_1^2  -  \theta_1^1  ) \partial_0+ \frac{1}{2} (i  \theta_1^2  + \theta_1^1) \partial_1 ~,
\end{align}
and
\begin{align}
    (D_0^1)^2 =  (D_0^2)^2 &=i  \partial_0 ~,  
& (D_1^1)^2 = (D_1^2)^2 &=i  \partial_1 ~.
\end{align}
 Under integration,
\begin{equation}
S= -  \frac{1}{2} \int d \sigma d \tau d \theta_1^1 d \theta_1^2 \; D_0^1  D_0^2 K |_{\theta_0^1=\theta_0^2=0} ~.
\end{equation}
Now,
\begin{equation}
D_0^1  D_0^2 K = K_{\mu \nu} D_0^1 \Phi^\mu D_0^2 \Phi^\nu +  K_{\mu}  D_0^1 D_0^2 \Phi^\mu ~.
\end{equation}
Due to \eqref{eq:chiralrelations}, we have
\begin{align}
 D_0^2 \Phi^{\alpha} &= - i  D_0^1 \Phi^{\alpha}  ~, &  D_0^2 \Phi^{\bar \alpha} &= + i  D_0^1 \Phi^{\bar \alpha} ~,
\end{align}
and 
\begin{equation}
\begin{split}
K_{\mu}  D_0^1 D_0^2 \Phi^\mu &= -i K_{\alpha}  D_0^1 D_0^1 \Phi^\alpha  + i K_{\bar\alpha}  D_0^1 D_0^1 \Phi^{\bar \alpha} \\
&=   K_{\alpha}  \partial_0 \Phi^\alpha  -  K_{\bar\alpha}  \partial_0  \Phi^{\bar \alpha} \\
&= 2 K_{\alpha}  \partial_0 \Phi^\alpha  + \text{total derivative.}
\end{split}
\end{equation}
Also,
\begin{equation}
\begin{split}
K_{\mu \nu} D_0^1 \Phi^\mu D_0^2 \Phi^\nu &= K_{\mu \alpha} D_0^1 \Phi^\mu D_0^2 \Phi^ \alpha + K_{\mu \bar \alpha} D_0^1 \Phi^\mu D_0^2 \Phi^{\bar\alpha}  \\
&= - i K_{\bar \beta \alpha} D_0^1 \Phi^{\bar \beta} D_0^1 \Phi^ \alpha + i K_{\beta \bar \alpha} D_0^1 \Phi^\beta D_0^1 \Phi^{\bar\alpha} \\
&= - 2 i K_{\bar \beta \alpha} D_0^1 \Phi^{\bar \beta} D_0^1 \Phi^ \alpha ~.
\end{split}
\end{equation}
Using  \eqref{eq:chiralrelations} again, we note that
\begin{align}
 D_0^1 \Phi^{\alpha} &=\tfrac{1}{\sqrt{2}} (\Dp^1 + \Dm^1) \Phi^{\alpha} =\tfrac{1}{\sqrt{2}}  (\Dp^1 + i \Dm^2) \Phi^{\alpha} = D_1^1 \Phi^{\alpha} ~, \\
 D_0^1 \Phi^{\bar \alpha} &=\tfrac{1}{\sqrt{2}} (\Dp^1 + \Dm^1) \Phi^{\bar \alpha} =\tfrac{1}{\sqrt{2}}  (-i \Dp^2 + \Dm^1) \Phi^{\bar\alpha} =-i  D_1^2 \Phi^{\bar\alpha} ~,
 \end{align}
so,
\begin{equation}
K_{\mu \nu} D_0^1 \Phi^\mu D_0^2 \Phi^\nu =  - 2 K_{\bar \beta \alpha} D_1^2 \Phi^{\bar \beta} D_1^1 \Phi^ \alpha ~,
\end{equation}
and
\begin{equation}
S= \int d^2 \sigma d \theta_1^1 d \theta_1^2 \; \left(  K_{\bar \beta \alpha} D_1^2 \Phi^{\bar \beta} D_1^1 \Phi^ \alpha   - K_{\alpha}  \partial_0 \Phi^\alpha    \right) \Big |_{\theta_0^1=\theta_0^2=0} ~.
\end{equation}
Denote $\theta^1\equiv \sqrt{i} \theta^1_1$,  $\theta^2 \equiv  \sqrt{i} \theta^2_1$ and $\partial \equiv  \partial_1$. Let  
\begin{align} \label{eq:defofD1D2}
D_1 & \equiv  - i \sqrt{i} D_1^1  |_{\theta_0^1=\theta_0^2=0} =  \frac{\partial}{\partial \theta^1} + \theta^1\partial ~,\\
D_2 & \equiv  - i \sqrt{i} D_1^2  |_{\theta_0^1=\theta_0^2=0}= \frac{\partial}{\partial \theta^2} + \theta^2\partial ~,
\end{align}
and $\phi^\mu \equiv  \Phi^\mu  |_{\theta_0^1=\theta_0^2=0}$. 
Then
\begin{equation} \label{eq:actioninhamiltonianform}
\begin{split}
S&= \int d^2 \sigma d \theta^2 d \theta^1 \; \left( i K_{\alpha}  \partial_0 \phi^\alpha  - K_{\bar \beta \alpha} D_2 \phi^{\bar \beta} D_1 \phi^ \alpha   \right)  \\
&= \int d^2 \sigma d \theta^2 d \theta^1 \; \left( i K_{\alpha}  \partial_0 \phi^\alpha  - \frac{1}{2} \mathcal{H} \right) ~,
\end{split}
\end{equation}
with 
\begin{equation}\label{eq:hamiltonian}
\mathcal{H}= K_{\bar \beta \alpha} D_2 \phi^{\bar \beta} D_1 \phi^ \alpha -  K_{\bar \beta \alpha} D_1 \phi^{\bar \beta} D_2 \phi^ \alpha ~.
\end{equation}

From \eqref{eq:actioninhamiltonianform} we see that the Hamiltonian density is given by \eqref{eq:hamiltonian}. The momenta is
\begin{align} \label{eq:defmomenta}
p_\alpha &= i K_{\alpha} ~, & p_{\bar \alpha} &= 0 ~.
\end{align}
The definitions of the momentas give the second class constraints $p_\alpha - i K_{\alpha} = 0$ and  $ p_{\bar \alpha} = 0$, leading to the Dirac brackets
\begin{align} \label{eq:Diracbrackets}
\{ \phi^\alpha ,  \phi^{\bar \beta} \}^* &= \omega^{\alpha \bar \beta} ~,&
\{ \phi^\alpha , p_\beta \}^* &= \delta_\beta^\alpha ~,&
\{ \phi^{\bar\alpha} , p_\beta \}^* &= i  \omega^{\bar \alpha \alpha} K_{ \alpha \beta} ~,
\end{align}
with the remaining brackets being zero.

Let us define new combinations of the derivatives $D_1$ and $D_2$: 
\begin{equation}\label{eq:Dpmcl}
D_\pm \equiv \frac{1}{\sqrt{2}} (D_1 \mp i D_2) ~.
\end{equation}
We can then write the Hamiltonian \eqref{eq:hamiltonian} as
\begin{equation}
\mathcal{H}= \G^c_-  - \G^c_+ ~,
\end{equation}
with 
\begin{equation}\label{eq:Gc}
\G^c_\pm= D_\mp p_\alpha  D_\pm \phi^\alpha + i K_{\alpha \beta} D_\pm \phi^\alpha D_\mp \phi^\beta  ~.
\end{equation}

\bibliographystyle{utphys}
\bibliography{articles}

\providecommand{\href}[2]{#2}\begingroup\raggedright\begin{thebibliography}{10}

\bibitem{Malikov1999}
F.~Malikov, V.~Schechtman, and A.~Vaintrob, ``{Chiral de Rham Complex},''
  \href{http://dx.doi.org/10.1007/s002200050653}{{\em Communications in
  Mathematical Physics} {\bfseries 204} no.~2, (July, 1999) 439--473},
  \href{http://arxiv.org/abs/math/9803041}{{\ttfamily arXiv:math/9803041}}.

\bibitem{Ben-Zvi2008}
D.~Ben-Zvi, R.~Heluani, and M.~Szczesnya, ``{Supersymmetry of the chiral de
  Rham complex},'' \href{http://dx.doi.org/10.1112/S0010437X07003223}{{\em
  Compositio Mathematica} {\bfseries 144} no.~02, (Feb., 2008) 503--521}.

\bibitem{Heluani2008a}
R.~Heluani, ``{Supersymmetry of the chiral de Rham complex II: Commuting
  Sectors},'' {\em {Int. Math. Res. Not. IMRN}} no.~{6}, (June, {2009})
  {953--987}, \href{http://arxiv.org/abs/0806.1021}{{\ttfamily arXiv:0806.1021
  [math.QA]}}.

\bibitem{Heluani:2008hw}
R.~Heluani and M.~Zabzine, ``{Generalized Calabi-Yau manifolds and the chiral
  de Rham complex},'' \href{http://dx.doi.org/10.1016/j.aim.2009.10.007}{{\em
  Adv. Math.} {\bfseries 223} (2010) 1815--1844},
\href{http://arxiv.org/abs/0812.4855}{{\ttfamily arXiv:0812.4855 [math.QA]}}.

\bibitem{Ekstrand2010a}
J.~Ekstrand, R.~Heluani, J.~K\"{a}ll\'{e}n, and M.~Zabzine, ``{Chiral de Rham
  complex on special holonomy manifolds},''
  \href{http://arxiv.org/abs/1003.4388}{{\ttfamily arXiv:1003.4388 [hep-th]}}.

\bibitem{malikov-lagrangian}
F.~Malikov, ``Lagrangian approach to sheaves of vertex algebras,''
  \href{http://dx.doi.org/10.1007/s00220-007-0403-3}{{\em Comm. Math. Phys.}
  {\bfseries 278} no.~2, (2008) 487--548}.

\bibitem{Ekstrand2009c}
J.~Ekstrand, R.~Heluani, J.~K\"{a}ll\'{e}n, and M.~Zabzine, ``{Non-linear sigma
  models via the chiral de Rham complex},'' {\em Advances in Theoretical and
  Mathematical Physics} {\bfseries 13} no.~4, (Aug., 2009) 1221--1254,
  \href{http://arxiv.org/abs/0905.4447}{{\ttfamily arXiv:0905.4447 [hep-th]}}.

\bibitem{kapranov}
M.~Kapranov and E.~Vasserot, ``Vertex algebras and the formal loop space,''
  \href{http://dx.doi.org/10.1007/s10240-004-0023-9}{{\em Publ. Math. Inst.
  Hautes \'Etudes Sci.} no.~100, (2004) 209--269}.

\bibitem{Kac1998}
V.~G. Kac, {\em {Vertex algebras for beginners}}, vol.~10 of {\em University
  lecture series}.
\newblock American Mathematical Society, second~ed., 1998.

\bibitem{Heluani2007}
R.~Heluani and V.~G. Kac, ``{Supersymmetric Vertex Algebras},''
  \href{http://dx.doi.org/10.1007/s00220-006-0173-3}{{\em Communications in
  Mathematical Physics} {\bfseries 271} no.~1, (Jan., 2007) 103--178}.

\bibitem{linshaw}
B.~H. Lian and A.~R. Linshaw, ``Chiral equivariant cohomology. {I},''
  \href{http://dx.doi.org/10.1016/j.aim.2006.04.008}{{\em Adv. Math.}
  {\bfseries 209} no.~1, (2007) 99--161}.

\bibitem{vaisman1994lectures}
I.~Vaisman, {\em {Lectures on the geometry of Poisson manifolds}}.
\newblock Progress in mathematics. Birkh{\"a}user Verlag, 1994.

\bibitem{Zabzine2006a}
M.~Zabzine, ``{Hamiltonian Perspective on Generalized Complex Structure},''
  \href{http://dx.doi.org/10.1007/s00220-005-1512-5}{{\em Communications in
  Mathematical Physics} {\bfseries 263} no.~3, (Jan., 2006) 711--722}.

\bibitem{Bredthauer:2006hf}
A.~Bredthauer, U.~Lindstr\"om, J.~Persson, and M.~Zabzine, ``{Generalized
  K\"ahler geometry from supersymmetric sigma models},''
  \href{http://dx.doi.org/10.1007/s11005-006-0099-x}{{\em Lett. Math. Phys.}
  {\bfseries 77} (2006) 291--308},
\href{http://arxiv.org/abs/hep-th/0603130}{{\ttfamily arXiv:hep-th/0603130}}.

\bibitem{MR2003030}
K.~Hori, S.~Katz, A.~Klemm, R.~Pandharipande, R.~Thomas, C.~Vafa, R.~Vakil, and
  E.~Zaslow, {\em Mirror symmetry}, vol.~1 of {\em Clay Mathematics
  Monographs}.
\newblock American Mathematical Society, Providence, RI, 2003.

\bibitem{Lindstrom:2005zr}
U.~Lindstr\"om, M.~Ro\v{c}ek, R.~von Unge, and M.~Zabzine, ``{Generalized
  K\"ahler manifolds and off-shell supersymmetry},''
  \href{http://dx.doi.org/10.1007/s00220-006-0149-3}{{\em Commun. Math. Phys.}
  {\bfseries 269} (2007) 833--849},
\href{http://arxiv.org/abs/hep-th/0512164}{{\ttfamily arXiv:hep-th/0512164}}.

\end{thebibliography}\endgroup

\end{document}